\documentclass[10pt,journal,compsoc]{IEEEtran}
\usepackage[nocompress]{cite}
 
\usepackage{cite}
\usepackage{amsmath,amssymb,amsfonts}
\usepackage{mathtools, nccmath}
\usepackage{algorithmic}
\usepackage{textcomp}
\usepackage[pdftex]{graphicx}
\usepackage{times,epsfig,amsmath}
\usepackage {verbatim}
\usepackage{ragged2e}
\usepackage{subcaption}
\usepackage{caption,setspace}
\usepackage[lined,commentsnumbered]{algorithm2e} 
\usepackage{url} 
\usepackage{xcolor} 

\makeatletter 
\newcommand{\mathleft}{\@fleqntrue\@mathmargin0pt} 
\newcommand{\mathcenter}{\@fleqnfalse} 
\makeatother 
\usepackage{authblk}
\begin{document}

\title{Cross Layer Resource Allocation in H-CRAN with Spectrum and Energy Cooperation}

\author{ Nazanin~Moosavi, Mahnaz~Sinaie, Paeiz~Azmi, Pin-Hsun~Lin and Eduard~Jorswieck 
  \thanks {\textbullet ~~ Nazanin Moosavi and Paeiz Azmi are with the Department of Electrical Engineering, Tarbiat Modarres University, Tehran, Iran, (e.mail : \{n\_moosavi,pazmi\}@modares.ac.ir)}
\thanks{\textbullet ~~ Mahnaz Sinaie is with the VTT Technical Research Centre of Finland, Espoo, Finland,(e.mail mahnaz.sinaie@vtt.fi)}
\thanks {\textbullet ~~ Pin-Hsun Lin and Eduard Jorswieck are with the TU Braunschweig , Germany, Braunschweig,(e.mail \{lin,jorswieck\}@ifn.ing.tu-bs.de)}
\thanks{\textbullet ~~The work of Mahnaz Sinaie was supported by Iran National Science Foundation (INSF) under grant 96007791.}
\thanks{\textbullet ~~~ The work of Paeiz Azmi was supported by Iran National Science Foundation (INSF) under grant 91059145.}
\thanks{\textbullet ~~~ The work of Pin-Hsun Lin was supported in part by DFG LI 2886/2-1 and Fast Secure.}
\thanks{\textbullet ~~~ The work of E. Jorswieck was partly supported by the German Research Foundation (DFG) under grant JO 801/24-1.}}

\IEEEtitleabstractindextext{%
\justify
\begin{abstract}
5G and beyond wireless networks are the upcoming evolution for the current cellular networks to provide the essential requirement of future demands such as high data rate, low energy consumption, and low latency to provide seamless communication for the emerging applications. Heterogeneous cloud radio access network (H-CRAN) is envisioned as a new trend of 5G that uses the advantages of heterogeneous and cloud radio access networks to enhance both the spectral and energy efficiency. In this paper, building on the notion of effective capacity (EC), we propose a framework in non-orthogonal multiple access (NOMA)-based H-CRAN to meet these demands simultaneously. Our proposed approach is to maximize the effective energy efficiency (EEE) while considering spectrum and power cooperation between macro base station (MBS) and radio remote heads (RRHs). To solve the formulated problem and to make it more tractable, we transform the original problem into an equivalent subtractive form via Dinkelbach algorithm. Afterwards, the combinational framework of distributed stable matching and successive convex algorithm (SCA) is then adopted to obtain the solution of the equivalent problem. Hereby, we propose an efficient resource allocation scheme to maximize energy efficiency while maintaining the delay quality of service (QoS) requirements for the all users. The simulation results show that the proposed algorithm can provide a non-trivial trade-off between delay and energy efficiency in NOMA H-CRAN systems in terms of EC and EEE and the spectrum and power cooperation improves EEE of the proposed network. Moreover, our proposed solution complexity is much lower than the optimal solution and it suffers a very limited gap compared to the optimal method.
\end{abstract}

}
\maketitle

\IEEEpeerreviewmaketitle

\section{Introduction}
\label{sec:introduction}
\justifying
\IEEEPARstart{T}{he} tremendous growth of wireless network services with a massive number of users incurs an urgency for designing energy efficient communication systems while guaranteeing quality of service (QoS) \cite{15}. Compared with the 4G communication networks, the 5G networks are required to achieve 1,000 times higher system capacity, at least 100 times higher energy efficiency, 50 times improvement in latency, and 100 times higher in connectivity density \cite{11}. As such, new techniques such as non-orthogonal multiple access (NOMA) and spectrum and energy cooperation in new network architectures such as heterogeneous cloud radio access networks (H-CRANs) have been envisioned as promising solutions to enable efficient 5G communication networks \cite{zhang2018energy}, \cite{QW1}.\par 
H-CRANs are such networks that can take the full advantage of both heterogeneous networks and cloud radio access networks simultaneously \cite{peng2014energy}. The H-CRAN architecture consists of a baseband unit (BBU) pool, fiber links, one macro base station (MBS), and some remote radio heads (RRHs). In this network, the centralized signal processing is performed in the BBU pool instead of distributed base stations (BSs) in the heterogeneous networks. The control plane and data plane are decoupled and the delivery of control and broadcast signalling is shifted from RRHs to MBS. Hence,  signaling overheads is reduced and the cooperation between different RRHs is permitted because of the centralized signal processing. \par
Another promising approach to improve the energy efficiency (EE), spectral efficiency (SE), massive connectivity, and transmission latency is using new multiple access transmission technique such as NOMA \cite{zhang2018energy},\cite{rbn}. In comparison with the conventional orthogonal multiple access (OMA) schemes, NOMA schemes use non-orthogonal resources to provide services for multiple users with the main advantage to increase fairness between them, while increasing the user's receiver complexity and multiple access interference. In particular, successive interference cancellation (SIC) is adjusted at receivers to mitigate the mutual interference caused by the non-orthogonal transmission. In order to get the full advantages of both H-CRAN architecture and NOMA technique simultaneously, we consider NOMA based H-CRAN systems.\par 
Another approach to improve system performance and EE in H-CRAN system is to apply spectrum and energy cooperation \cite{QW1} -\cite{liu2013green}. In \cite{QW1}, there was a cooperation strategy in small cell networks where the small cell base station serves the macro cell users while it can use some bandwidth of the macro cell\cite{QW1}. Moreover, in \cite{guo2014joint}, a joint energy and spectrum cooperation scheme between different cellular systems was considered in order to decrease their operational costs. In\cite{xie2012energy}, energy efficient resource allocation for heterogeneous cognitive radio networks with femtocells were studied, where a cognitive BS maximizes its energy efficiency by allocating the spectrum borrowed from the primary networks to the femtocells. In \cite{liu2013green}, the authors considered spectrum sharing in a green communication system to reduce energy consumption as much as possible with guaranteed users QoS. Hence, we will apply the spectrum and energy cooperation in our NOMA based H-CRAN system to enhance energy efficiency.\par
As it was mentioned, one of the main goals for 5G network design is to provide very short end-to-end communication delay. In communication networks with delay sensitive applications, optimizing solely the physical layer is not sufficient to guarantee the QoS requirements due to the random variations of wireless channel conditions. As an efficient alternative, one should consider the delay requirement of each user statistically \cite{15, cheng2014heterogeneous, cheng2016decentralized, amjad2019effective}. Hence, the notion of effective capacity was proposed in \cite{WU1} which relates the Shannon capacity to delay QoS requirements by introducing the data link layer characteristics into the communication theory. There are some references considering the statistical delay in CRAN architecture using effective capacity notion. For example, in \cite{chen1}, the authors developed an effective content RRH clustering and caching scheme to reduce the interference and offload the traffic of the backhaul and fronthaul. In \cite{zhao1}, the authors considered effective capacity of the cluster content cashing to study the impacts on the energy efficiency and delay. In \cite{Du1}, the authors studied the distributed wireless MIMO links and minimized the BS usages while considering heterogeneous statistical delay QoS constraints. In \cite{ismaiel2018analysis}, the authors establish a Markovian model to derive a closed-form expression of effective capacity in device to device (D2D) communication in terms of QoS and the transmission rate. In the context of EC in NOMA based networks, in \cite{choi2017effective}, the power control policy is applied to meet the delay QoS constraints in a NOMA based network using the partial effective capacity. The authors in \cite{ns} obtain the EC of NOMA based virtualized wireless network. In \cite{2}, the achievable EC under the per-user statistical delay QoS requirements, for a downlink NOMA based network is obtained. However, the analysis of the statistical delay and energy efficiency simultaneously for the cross layer resource allocation in the NOMA based H-CRAN networks have been missed in the literature, which, given the discussion above, appears an important setting to investigate.\par
In this work, we obtain the EC and EEE definition of the NOMA based H-CRAN systems while considering spectrum and energy cooperation between MBS and RRHs. Afterward, we formulate the corresponding resource allocation problems to allocate power, RRHs, and bandwidth to achieve delay aware energy efficient framework for the proposed H-CRAN scheme. The objective function is EEE which is the ratio of the effective capacity to the total power consumption where the total power consumption has three parts e.g. fiber link power consumption, active RRH power consumption, and the circuit power consumption. The considered EEE optimization problem is a non-convex, NP-hard, intractable, and fractional programming problem. Therefore, we apply Dinkelbach algorithm with stable matching framework and successive convex approximation (SCA) to allocate the resources efficiently.  The  simulation results demonstrate that the spectrum and energy cooperation improves EEE compared to the other networks. Moreover, EC is a decreasing function of delay and there is still a trade off between delay and effective energy efficiency. In addition, our proposed approach has lower complexity than the optimal solution with a negligible gap with the optimal results.

The remainder of this paper is organized as follows. Section II describes our system model and problem statement. In Section III, the proposed solution to the problem is provided. The performance of the proposed algorithm and our system model through different numerical specifications are examined in Section IV. Finally, Section V concludes this paper. 
\begin{table}
\footnotesize
\caption{ List of variables}
\label{table}
\setlength{\tabcolsep}{3pt}
\begin{tabular}{|p{1.5cm}|p{6cm}|}
\hline
Variables&
Definitions\\
\hline
\hline
$K$& Number of users connected to the RRHs \\ 
$Z$& Number of users connected to MBS\\ 
$B$ & Bandwidth of each RRHs\\ 
$W_{MC}$ & Total bandwidth of MBS\\ 
$W_z$ & Allocated bandwidth to the $z$-th MBS user\\ 
$M$ & Number of RRHs\\ 
$\theta$ & Delay QoS exponent\\ 
$\mathbb{E\{ . \}}$& Expectation\\ 
$T_f$& Whole time slot duration\\ 
$g_{k,m}$ & Channel  gain between the $k$-th user and the $m$-th RRH \\ 
$p_{z}$& Allocated power from MBS to the $z$-th user \\ 
$p_{k,m}$ & Allocated power from the $m$-th RRH to the $k$-th user \\ 
$x_{z,m}$& MBS user selection strategy by RRHs\\ 
$q_{k,m,z}$& Allocated power by the $m$-th RRH to the $k$-th user on the bandwidth $W_z$\\ 
$b_{m,z}$ &  Allocated bandwidth from $W_z$ to the users of the $m$-th RRH \\ 
$h_{k,m,z}$&  Channel gain between the $k$-th user and the $m$-th RRH on the bandwidth $W_z$\\ 
$w_{m,z}$&  Allocated bandwidth from $W_z$ of the $m$-th RRH to the $z$-th MBS user  \\ 
$Q_{m,z}$ &   Allocated power to the $z$-th MBS user by the $m$-th RRH\\ 
$H_{m,z}$ &   Channel power between the $z$-th MBS user and the $m$-th RRH\\ 
$\zeta_m$ & Reciprocal of drain efficiency of the $m$-th RRH\\ 
$P^m_{max}$ & Maximum available power of the $m$-th RRH\\ 
$E_c$ & Effective capacity\\ 
$\eta_{EEE}$ & Effective energy efficiency\\ 
$ P_{c,m}$ & Circuit power consumption of the $m$-th RRH\\ 
$ P_{m,f}$ &  Power consumption the $m$-th fronthaul fiber link\\ 
$R^{\text{ave-max}}_m$& Average rate of the $m$-th fronthaul link\\
$R^{\text{ave-max}}_z$ & Average required rate of the $z$-th MBS user\\
\hline
\end{tabular}
\label{tab1}
\end{table}
\section{System Model }\label{system model} 
\begin{figure}
   \includegraphics[width=8.3 cm , height=4.7 cm]{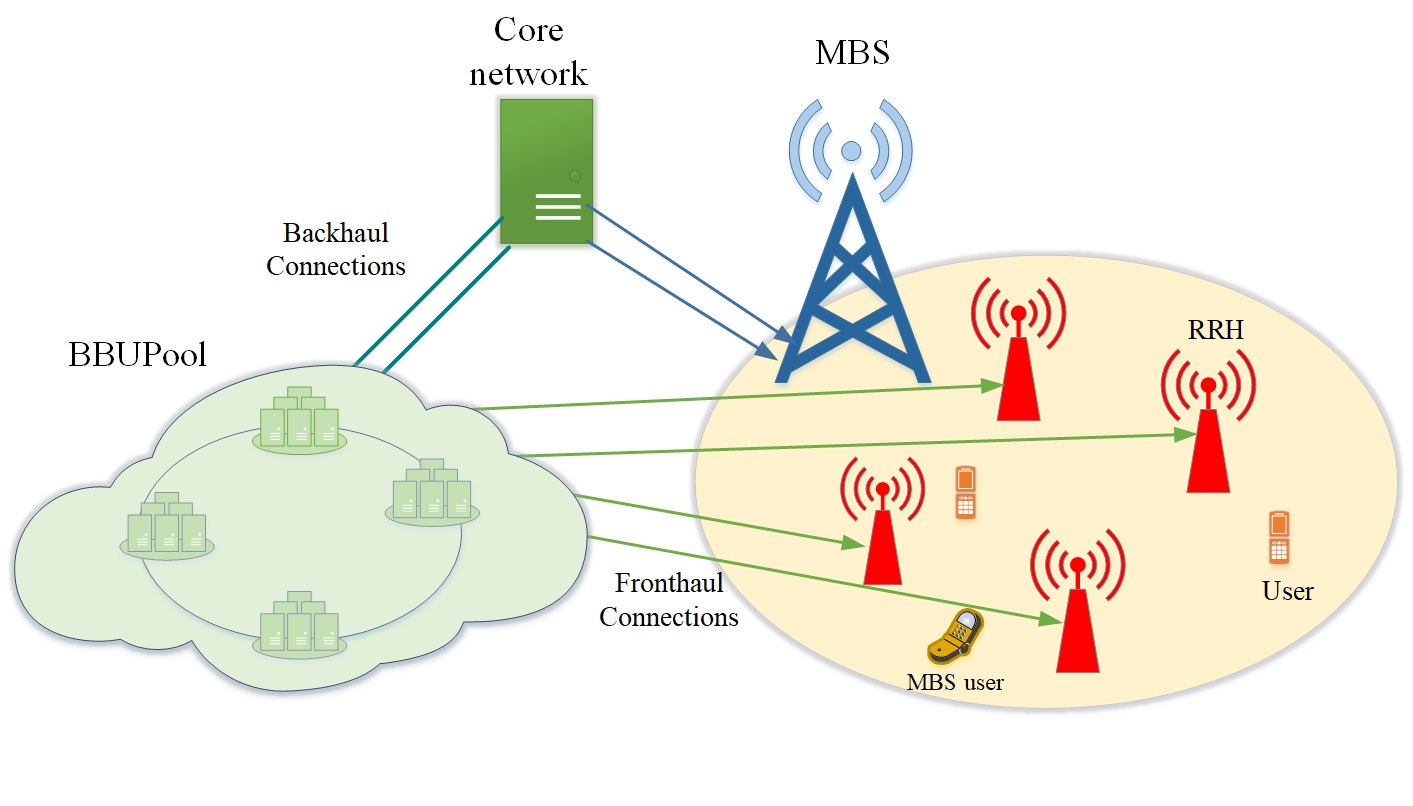} 
    \caption{\footnotesize{H-CRAN system with one macro base station (MBS), some RRHs, MBS users and the other users.} }
    \vspace{-0.5cm} 
    \label{fig:sys} 
\end{figure}

We study a two-tier downlink transmission in a H-CRAN network consisting of one MBS serving $Z$ users and $M$ RRHs serving $K$ users, as shown in Fig.\ref{fig:sys}. Let $\mathcal{M}=\{1,2,...,M\}$ denote the set of RRHs,  $\mathcal{Z}=\{1,2,...,Z\}$ denote set of MBS users and $\mathcal{K}=\{1,2,...,K\}$ denote set of users supported by RRHs. The MBS  bandwidth is $W_{MC}$ and the available total bandwidth for all RRHs is $W$ which is equally divided into $M$ sections with a dedicated bandwidth $B$ for each RRH. We further consider that all the RRHs encode their received data using NOMA technique and then transmit to the users in the downlink. We study the practical situation where there is spectrum and energy cooperation between RRHs and MBS. Channels between the users and the RRHs are assumed to be flat fading. Table I summarizes the parameters and symbols used in the system model.\par
The fronthaul capacity for each RRH is practically limited, in general it can only receive the data for a selected set of users from the BBU pool, and then forward them to the selected users in the NOMA-based downlink transmission. As a result, each RRH $m$ transmits only on the corresponding set of users whose data are received from the BBU pool. 
Hence, on the system model assumptions, the received signal at the $k$-th user is given by
\mathleft
\begin{equation}
\mathbf{y}_{k}=  \mathbf{x}_{k}  \mathbf{g}_{k} +\! \sum_{k'\neq{k}}\!\!  \mathbf{x}_{k'} \mathbf{g}_{k}+\mathbf{z},
\end{equation}
where $\mathbf{x}_{k}\!\!\!=\!\!\![x_{k,1},... ,x_{k,M}]\!=\![\sqrt{p_{k,1}},... ,\sqrt{p_{k,M}}]\!\!$ is the  transmitted signal to the $k$-th user, $p_{k,m}$  is the allocated power from the $m$-th RRH to the $k$-th user, $\mathbf{g}_{k}=[g_{k,1},g_{k,2},...,g_{k,M}]$ is the channel coefficient and $\mathbf{z}$ denotes white Gaussian noise having zero mean and variance $\sigma^2$.

In order to decode signals, in the NOMA-based system, the SIC process is implemented at each user receiver.  To get the desired signal, each SIC receiver first decodes the dominant interference which are sufficiently stronger than the receiver’s desired signal and then subtracts them from the superposed signal\cite{concept}. Hence, the SINR condition is satisfied in the following Lemma.\\  
\emph{\textbf{Lemma 1:}} The SINR condition $|g_{k',m}|^2\leq |g_{k,m}|^2$ is satisfied automatically in our considered NOMA system.
\begin{IEEEproof}  The proof is provided in Appendix A. \end{IEEEproof}

The SINR of the $k$-th user which is supported by the $m$-th RRH after performing SIC is as follow
\begin{equation} 
\gamma_{k,m}=\frac{p_{k,m} {|g_{k,m}|}^2}{\sum_{\substack{\footnotesize{ k'\neq{k}}\\ {|g_{k'\! ,m}|}^2\leq {|g_{k,m}|}^2}}p_{k',m} |g_{k,m}|^2+\footnotesize{B N_0}}. 
\end{equation}
Therefore, the achievable data rate is $r_{k,m}=B T_f \log_2(1+ \gamma_{k,m})$ where $T_f$ is time frame duration. Moreover, it should be noticed that each user can only be connected to one RRH and cooperation between RRHs due to the central signal processing is restricted. The exclusive RRH allocation strategy can be written as 
\begin{align}
    p_{k,m}p_{k,m'}=0, \forall m\neq m', m,m' \in \mathcal{M}.
\end{align}
 This statement ensures that each user can be at most connected to one RRH. Then, if $p_{k,m}\neq 0$ for $m$-th RRH then $p_{k,m'}=0$ for any RRH $m\neq m'$ \cite{alpha1}.
 
 \textbf{Remark 1:}
\textit{It is worth noting that H-CRAN network decouples the control and data plane which results in the significant signaling overhead saving in the radio connection between RRHs and BBU \cite{sig}. It is because instead of interaction between BBU pool and each RRH for interference management, the BBU pool is interfaced with the MBS for inter-tier interference coordination and totally the MBS is configured with the entire communication functionalities from physical to network layers, and therefore the delivery of control and broadcast signalling is shifted from RRHs to MBS, which decreases the capacity and time delay constraints on the fronthaul. Hence, the signaling overhead and complexity  between BBU, RRHs and users are negligible and we do consider such overheads neither in the analytical nor the simulation analyses.}
 
\subsection{Spectrum and Energy Cooperation} 
In our system model, we consider a spectrum and energy cooperation scenario which consists of MBS and some RRHs. Each RRH can provide higher data rate by getting more bandwidth from MBS and the MBS aims at offloading the data traffic of some of its cell edge users to the RRH in order to decrease its own energy consumption with the spectrum and energy cooperation. By considering that, each MBS user can be served by only one RRH, the user selection strategy is denoted by $x_{z,m}$ as follows:  
\begin{equation}\label{MCS} 
 x_{z,m}=\left\{ \begin{array}{ll} 
  1&\text{If MBS user $z$ is served by RRH $m$} \\ 
  0&\text{otherwise.}\\ 
   \end{array} \right. 
\end{equation}

The $m$-th RRH which is selected to serve the $z$-th MBS user with a dedicated bandwidth $W_z$, allocates $w_{m,z}$ bandwidth to that MBS user. Hence, the  achievable data rate of $z$-th user over $m$-th RRH is $r_{m,z}=w_{m,z} T_f \log_2(1+\frac{Q_{m,z}|H_{m,z}|^2}{w_{m,z} N_0})$ where $Q_{m,z}$ is the allocated power to the $z$-th MBS user by the $m$-th RRH, and $H_{m,z}$ is the channel gain between the $z$-th MBS user and the $m$-th RRH on the bandwidth $w_{m,z}$. Then $b_{m,z}$ which is the remaining bandwidth of $W_z$,  is assigned to transmit data to the users connected to  the $m$-th RRH. Consequently,
 the achievable data rate of the $k$-th user connected to the $m$-th RRH on the assigned bandwidth $b_{m,z}$, is expressed as 
\mathleft
\begin{align}
r_{k,m,z}= b_{m,z} T_f \log_2 ( 1+\nu_{k,m,z}),
\end{align}
where  $\nu_{k,m,z}\!=\!\frac{q_{k,m,z} |h_{k,m,z}|^2}{b_{m,z} N_0+I_{k,m,z}}$. In this equation,  $I_{k,m,z}$ is $\sum_{\substack{k'\neq{k}\\ |h_{k',m,z}|^2 \leq |h_{k,m,z}|^2}}\!\!\! q_{k',m,z} |h_{k,m,z}|^2$ 
,$h_{k,m,z}$ denotes the channel gain between the $k$-th user and the $m$-th RRH on the bandwidth $b_{m,z}$ and $q_{k,m,z}$ is the corresponding transmit power respectively. \par
It is worth to stress that based on \emph{Lemma 1}, the SINR condition for successive decoding is satisfied automatically in our proposed scheme and also for $r_{k,m,z}$. Hence, the achievable total data rate of the $k$-th user by considering spectrum and energy cooperation is given as 
\begin{equation}\label{rate1}
r_{k_t}=\sum_{m=1}^M r_{k,m}+ \sum_{z=1}^Z \sum_{m=1}^M x_{z,m} r_{k,m,z}. 
\end{equation} 
\subsection{Effective Capacity}\label{ecpf} 
The theory of link-layer effective capacity was introduced in \cite{e1c} to take into account the communication delay, by incorporating the users' delay-QoS requirements into the Shannon capacity. The effective capacity determines the maximum arrival rate that the channel can support with the given delay exponent $\theta$. It captures the delay QoS constraints including link layer characteristics such as the queue distribution and the buffer overflow probability \cite{15}. Assuming independent and identically distributed (i.i.d.) block fading, this leads to express the effective capacity as \cite{WU1} and \cite{tang1}
\begin{equation}\label{effectiveC}
E_c= -\frac{1}{\theta}\ln\left(\mathbb{E}_{\gamma}\{e^{-\theta  r}\}\right),
\end{equation}
where $\mathbb{E}$ is the statistical expectation concerning the so-called service rate $r=\log_2\left(1+\text{SINR}\right)$, with SINR which is modeled as a random variable due to the fading channel $\gamma$. A larger $\theta$ enforces more stringent delay-QoS constraints\cite{wu2005utilizing}, whereas if $\theta=0$ no delay QoS is enforced, and the effective capacity can be shown to simplify the usual Shannon capacity.
Thus, the effective capacity can be seen to be a generalization of the Shannon capacity, which jointly accounts for both the communication reliability and delay. Instead, the Shannon capacity considers only the reliability aspect, without explicitly accounting for the communication delay.

 Moreover, effective capacity maximization is clearly technically more challenging than maximizing the Shannon capacity, mainly due to the expectation operator in \eqref{effectiveC} and  the more involved functional form.
Hence, Plugging (\ref{rate1}) into \eqref{effectiveC} yields the effective capacity of $k$-th user in $\mathcal{K}$, namely
\begin{align} \label{f1} 
&E^{k}_{c}\!\!=\!\!\! \sum_{m=1}^M{\frac{-1 } {\theta_k}} \log \mathbb{E}_{g}\{ e^{-\theta_k B T_f \log_2(1+\gamma_{k,m}) }\} \nonumber \\ 
&\hspace{1.5em} \!\!+\!\!\ \sum_{z=1}^Z \! \sum_{m=1}^M \frac{-x_{z,m}\! }{\theta_k} \log \mathbb{E}_{h}\!  \{ e^{-\theta_k  b_{m,z} T_f\! \log_2(1+\nu_{k,m,z})\!}\! \},
\end{align} 
where $\mathbb{E}_{g}$ and $\mathbb{E}_{h}$ give the statistical average of inner arguments with respect to channel gains $g$ and $h$ respectively.
\subsection{power consumption model}
Here, we turn our attention to the derivation of the power consumption $P_T$ in the considered scenario. By definition, the effective capacity considers a time frame which embraces multiple channel realizations, due to the expectation in \eqref{effectiveC}. Accordingly, the consumed power must be evaluated over the same time frame. Our power consumption model consists of two parts, dynamic and static power consumption. Dynamic power consumption includes both the transmit power consumption for MBS users and other users connected to RRHs as follow:\\
\begin{align}
& P_d =\mathbb{E}_{g}\left \{\sum_{m=1}^M \sum_{k=1}^k  \zeta_m  p_{k,m}\right \}
 \nonumber\\ 
&\hspace{1.2em}+\mathbb{E}_{h}\left \{\sum_{z=1}^Z \sum_{m=1}^M \sum_{k=1}^K  \zeta_m  x_{z,m}  q_{k,m,z}\right \}\nonumber\\ 
&\hspace{1.2em}+ \mathbb{E}_{H}\left\{\sum_{z=1}^Z \sum_{m=1}^M \zeta_m x_{z,m}Q_{z,m}\right\}, 
\end{align}
where $\zeta_m$ is the reciprocal of drain efficiency of the $m$-th RRH and the value of it depends on the specific type of the RRH,  and $\mathbb{E}_{H}$ gives the statistical average of included arguments with respect to channel gains $H$.
The static power consumption consists of fiber power consumption, and the power consumed in the hardware blocks of the network nodes. Hence, the total power consumption is given as 
\begin{align} 
& P_T =\sum_{m=1}^M p_{c,m} +\sum_{m=1 }^M P_{m,f}+\mathbb{E}_{g}\left \{\sum_{m=1}^M \sum_{k=1}^k  \zeta_m p_{k,m}\right \} 
 \nonumber\\ 
&\hspace{1.2em}+\mathbb{E}_{h}\left \{\sum_{z=1}^Z \sum_{m=1}^M \sum_{k=1}^K  \zeta_m  x_{z,m} q_{k,m,z}\right \}\nonumber\\ 
&\hspace{1.2em}+ \mathbb{E}_{H}\left\{\sum_{z=1}^Z \sum_{m=1}^M \zeta_m x_{z,m}Q_{z,m}\right\}, 
\end{align} 
wherein $p_{c,m}$ is the circuit power consumption of the $m$-th RRH, and $P_{m,f}$ is the power consumption of fiber links.
\subsection{Problem formulation }\label{PF}
Our main goal is to maximize the system bit-per-Joule effective energy efficiency while guaranteeing throughput, energy efficiency and delay-QoS requirement simultaneously. It is defined as the ratio  
\begin{align}
\eta _{EEE}= \frac{E_c}{P_T},
 \end{align}
In this context, the radio resource allocation problem to be tackled is the maximization of the effective energy efficiency via jointly optimizing transmit powers, and bandwidth, $\{p_{k,m}$, $q_{k,m,z}$, $Q_{m,z}$, $b_{m,z}$, $\forall k\in \mathcal{K}, m\in \mathcal{M}, z\in \mathcal{Z}$\}, subject to QoS requirements, total power constraint and RRH and MBS user selection constraints. Formally, The system EEE maximization problem is given as

\null 

\begin{equation} \label{TS_EE} 
\mathcal{P}_1:\max_{\substack{\!\left\{p_{k,m}\right\},\!\left\{\! q_{k,m,z}\!\right\} \\ \!\{\! Q_{m,z}\},\{b_{m,z}\},\{x_{z,m}\}}} \eta _{EEE}= \frac{\sum_{k=1}^K E^{k}_{c}}{P_T}\\  
\end{equation} 

\text{subject to:}\\
\begin{equation} {
\begin{cases}
\text{C1:}~E^{k}_{c} \geq  R^{\text{ave-max}}_m,~\forall k ,m ,\nonumber\\
 \text{C2:}~ \mathbb{E}_{H}\!\{\! w_{z,m} T_f \log_2\!(1+\frac{Q_{m,z}H_{m,z}}{w_{m,z} N_0}\!)\!\!\}\!  
\geq\!  x_{z,m}R^{\text{ave-max}}_z \!, ~ \forall z  \nonumber \\ 
 \text{C3:}~ \sum_{m=1}^M b_{m,z} +w_{m,z} \leq   \sum_{m=1}^M x_{z,m} W_z, ~ \forall z \nonumber \\ 
 \text{C4:} ~\mathbb{E}_{g}\!\! \left\{\!\sum_{k=1}^K\! p_{k,m}\! \right\} +\mathbb{E}_{H} \!\left \{\sum_{z=1}^Z x_{z,m}Q_{m,z}\!\right\}\\
 + \mathbb{E}_h\! \! \left\{\!\sum_{z=1}^Z\!\sum_{k=1}^K \! x_{z,m} q_{k,m,z}\!\right\}
\leq P^m_{\text{max}},~ \forall m \nonumber\\ 
 \text{C5:}~ p_{k,m} \geq 0,~ \forall  k ,m,\nonumber\\ 
 \text{C6:} ~b_{m,z} , q_{k,m,z} \geq 0,~ \forall m,z, \nonumber\\ 
 \text{C7:} ~Q_{m,z}, w_{m,z} \geq 0, ~\forall m,z,\nonumber\\ 
  \text{C8:}~ p_{k,m} p_{k,m'}=0, ~\forall m,m'\in \mathcal{M}, k\in \mathcal{M},\nonumber\\
 \text{C9:} ~x_{z,m} \in \{0,1\}, ~\forall z,m ,\nonumber\\ 
 \text{C10:} ~\sum_{z=1}^Z x_{z,m} \leq 1 , ~\forall m,  \nonumber\\ 
  \text{C11:} ~\sum_{m=1}^M x_{z,m} \leq 1 , ~\forall z,  \nonumber\\ 
 \end{cases}}
\end{equation}
where C1 is to guarantee the statistical delay bound of the $k$-th user, C2 guarantees the average required rate for the $z$-th MBS user, C3 ensures that the bandwidth allocated to RRH users and  the $z$-th MBS user does not exceed the available bandwidth, $W_z$
that has been licensed to the $z$-th MBS user.  C4 ensures that the average transmit power by the $m$-th RRH is non-negative and below the maximum feasible average transmit power, $P^m_{\text{max}}$, C5, C6, and C7 ensure that all allocated powers and bandwidth are non-negative, C8 indicates whether the $k$-th user is supported by the $m$-th RRH, and guarantees that each user is not connected to more than one RRH. C9 indicates whether to serve the $z$-th MBS user or not and C10 and C11 is to ensure that each RRH can just served one MBS user and each MBS user is not served by more than one RRH.

The EEE maximization problem is non-concave mixed integer fractional programming and therefore direct use of standard convex optimization tool is not possible. To tackle this burden, we applied stable matching framework, SCA and Dinkelbach algorithm. Stable matching is one of the well known approach which is suited to analysis of mixed-integer assignment problem. It is implemented inside Dinkelbach algorithm to obtain the MBS user assignment. Moreover, SCA is also  considered to deal with the non-concavity of the numerator. Consequently, Dinkelbach algorithm solves the optimization problem with polynomial complexity when the numerator and denominator are respectively concave and convex functions over a convex set which is shown to converge with super-linear convergence rate\cite{42}. The proposed algorithm is applicable from the point of the complexity and it guarantees the convergence to the suboptimal solution. In the simulation results, we compare the suboptimal approach with the optimal approach which is branch and bound algorithm demonstrating that our proposed suboptimal solution has a negligible gap with the optimal solution.
\section{Proposed solution} 
After formulating the resource allocation problem in Section \ref{PF}, this section is devoted to the development of the corresponding radio resource allocation algorithms. The problem \eqref{TS_EE} is non-convex due to the integer variable $\{x_{z,m} \}_{z,m}$ and has a fractional form. Therefore, there is no generally efficient method to solve this problem and the complexity grows exponentially.\par

At first, we turn our attention to the constraints C2 and C3 of problem $\mathcal{P}_1$. There is a discussion about network with a spectrum and power cooperation as how much power should RRH use to serve an MBS user against how much bandwidth the MBS user can give to the RRH. Since it is always useful for RRH not only to seek as much as bandwidth but also consuming as less as power in the spectrum and energy cooperation with the MBS, we can consider equality at the optimal solution so the constraint C2 and C3 can be satisfied in equality and change to
\mathleft
\begin{align}
& C2:~ \mathbb{E}_{H}\!\{\! w_{z,m} T_f \log_2\!(1+\frac{Q_{m,z}H_{m,z}}{w_{m,z} N_0}\!)\}
= x_{z,m} R^{\text{ave-max}}_z \!  , ~ \forall z \\
& C3: ~ b_{m,z} + w_{m,z} = x_{z,m}W_z,  ~ \forall z,m, 
\end{align}
Hence, we have $Q_{m,z}= \mathbb{E}_{H}\{\frac{w_{m,z} N_0}{H_{m,z}} (2^{\frac{R_z^{\text{ave-max}}}{T_f w_{m,z}}}-1)\}, ~ \forall z$. Moreover, since $b_{m,z}=W_z-w_{m,z}$, the optimization variable $b_{m,z}$ is replaced by $w_{m,z}$ for simplicity. Then, we have $\eta _{EEE}=\frac{\sum_{k=1}^K \widehat{E}^{k}_{c}}{\widehat{P}_T}$ where $\widehat{E}_c$ and  $\widehat{P}_T$ are defined as follows
\begin{align} \label{f2} 
&\widehat{E}_{c}\!\!=\!\!\!\sum_{k=1}^K \sum_{m=1}^M{\frac{-1} {\theta_k}} \log \mathbb{E}_{g}\left\{ e^{-\theta_k B T_f \log_2(1+\gamma_{k,m}) }\right\} \nonumber \\ 
&\hspace{1.1em}+\sum_{k=1}^K \sum_{z=1}^Z \! \sum_{m=1}^M \frac{-x_{z,m}\! }{\theta_k} \log \mathbb{E}_{h}\!  \left\{ e^{-\theta_k  (W_z-w_{m,z}) T_f\! \log_2(1+\nu_{k,m,z})\!}\! \right\},
\end{align} 
\begin{align} 
&\!\widehat{P}_T=\! \! \sum_{m=1}^M p_{c,m}+\mathbb{E}_{g}\left \{\sum_{m=1}^M \sum_{k=1}^k  \zeta_m  p_{k,m}\right \}+\sum_{m=1 }^M P_{m,f}\nonumber\\ 
&\hspace{1.2em}+\mathbb{E}_{H}\left\{\sum_{z=1}^Z \sum_{m=1}^M \zeta_m x_{z,m} \frac{w_{m,z} N_0}{H_{m,z}} (2^{\frac{R_z^{\text{ave-max}}}{T_f w_{m,z}}}-1)\right\}\nonumber \\ 
&\hspace{1.2em}+\mathbb{E}_{h}\left\{\sum_{z=1}^Z \sum_{m=1}^M \sum_{k=1}^K  \zeta_m x_{z,m} q_{k,m,z}\right\}. 
\end{align}
Accordingly, we have the following problem instead of problem $\mathcal{P}_1$ which will be tackled in the following subsections. 
\mathleft
\begin{align} \label{EEprob} 
\mathcal{P}_2: \max_{\substack{\!\left\{p_{k,m}\!\right\}_{k,m},\!\left\{\! q_{k,m,z}\!\right\}_{k,m,z}\ , \!\left\{\! w_{m,z}, x_{z,m}\!\right\}_{m,z}}}  \eta _{EEE}
\end{align}
\text{subject to}
\begin{equation} {
\begin{cases}
\text{C1:}~ E^{k}_{c} \geq  R^{\text{ave-max}}_m,~\forall m, k ,\nonumber\\
 \text{C4:} ~\mathbb{E}_{g}\!\! \{\!\sum_{k=1}^K\!  p_{k,m}\! \} +\mathbb{E}_{H} \!\{\sum_{z=1}^Z x_{z,m} Q_{m,z}\!\} \\
 + \mathbb{E}_h \{\!\sum_{z=1}^Z\!\sum_{k=1}^K x_{z,m} \! q_{k,m,z}\!\}
\leq P^m_{\text{max}},~ \forall m, \nonumber\\ 
 \text{C5:}~ p_{k,m} \geq 0,~ \forall  m ,k,\nonumber\\ 
 \text{C6:} ~ q_{k,m,z} \geq 0,~ \forall m,z  ,\nonumber\\ 
 \text{C7:} ~ w_{m,z} \geq 0, ~\forall m,z ,\nonumber\\ 
\text{C8:} ~p_{k,m} p_{k,m'}=0, ~\forall m,m', k. \nonumber
 \end{cases} }
\end{equation} \vspace{\baselineskip}
\textbf{Remark 2:}
\textit{It is worth noting that it is proved in \cite{QW1} that C2 and C3 are satisfied in the equality. It is due to the fact that by using spectrum and energy cooperation although we focus on improving the effective energy efficiency of RRHs, the EEE of the MBS will also improve. Because MBS desires to offload its users with poor channel condition which it needs more power to be served. Therefore, the EEE of the MBS will obviously increase via offloading. Furthermore, at the optimal point the most energy efficient strategy for each RRH is to search as much as bandwidth while consuming less power. Hence, in order to maximize the achievable bandwidth, the constraint C3 should reach its maximum value which is $W_z$ and therefore constraint C3 changes to equality. In addition, each RRH desires to consume less power, which means the $z$-th MBS user rate should be in its minimum value and it happens when it reaches its average rate value $R^{\text{ave-max}}_z$, so the constraint C2 changes to equality. }
 
\subsection{Dinkelbach Algorithm} \label{dinkel}

The next step is to tackle the fractional programming and non-convexity of the constraints. The optimization problem $\mathcal{P}_2$ is a non-linear fractional problem, which can be solved with the fractional programming theory \cite{alessio1}. Based on the fractional programming, the problem is transformed into an equivalent subtractive form via the parametric approach.  An auxiliary function $:\mathbf{q} \in \mathbb{R} \longrightarrow F(\mathbf{q})$ is introduced and the following problem must be solved.
\begin{equation}\label{fracmod}
\mathcal{P}_3:F(\mathbf{q})=\text{max}_ {\left\{p_{k,m}, w_{m,z}, q_{k,m,z}, x_{z,m}\right\}} \widehat{E}_{c}-\mathbf{q} \widehat{P}_T, 
\end{equation}
 where $\mathbf{q}$ is a non-negative parameter. Hence, the optimal solutions of $\mathcal{P}_2$ and $\mathcal{P}_3$ are related as $\mathbf{q}^* = \eta^{*}_{EEE}$. Hence, we can conclude that, a pair of power and bandwidth allocation strategy is a global solution of $\eta_{EEE}$ if and only if $F(\mathbf{q}^*) = 0$, with $\mathbf{q}^*$ being the maximum value of $ \eta_{EEE}$, i.e, $\mathbf{q}^* =  \eta^{*}_{EEE}$. Hence, the maximization of  $\eta_{EEE}$ is the same as finding the zeros of the auxiliary function $F(\mathbf{q})$. This can be accomplished by the well-known Dinkelbach algorithm which is an iterative approach where a sequence of the equivalent subtractive form (\ref{fracmod}) is solved at each iteration of this algorithm to update the auxiliary variable $\mathbf{q}$. It can be concluded that solving $\mathcal{P}_2$ is resort to obtaining $\mathbf{q}$ with $F(\mathbf{q}^*) = 0$. 
 
Although the original problem $\mathcal{P}_2$ can be transformed into $\mathcal{P}_3$ with an equivalent solution, $\mathcal{P}_3$ is obvious a non-convex problem due to the non-integer variable and existence of co-channel interference terms in rate formulation. Hence, to make $\mathcal{P}_3$ tractable, a combinational framework of stable matching and SCA are introduced in the following sections. It should be stressed that Dinkelbach’s algorithm is guaranteed to converge to the global solution of the corresponding fractional problem, provided one is able to globally solve the inner problem in \eqref{fracmod}, regardless of any concavity / convexity property of the numerator and denominator of the fractional objective to maximize.\\

\subsection{Power and Bandwidth Allocation Algorithm }\label{scale}
Due to the non-convexity of problem \ref{fracmod}, a logarithmic approximation method based on SCA  is adapted with any given $\mathbf{q}$ and $\{x_{z,m}\}_{z,m}$. Let us define $\Psi_m$ which is a set of MBS users supported by each RRH, i.e., $\Psi_m=\{z | x_{z,m}= 1,  z \in \mathcal{Z}\}=\{\Psi_{1,m},...,\Psi_{Z,m} \}$. By considering a given $\Psi_m$, problem $\mathcal{P}_3$ is just a joint bandwidth and  power allocation problem but it still has non-convexity in the rate formulation.\\
The main idea of SCA method is to approximate a non-convex problem into a series of solvable problems, to obtain the near optimal solution satisfying KKT conditions of the original problem. It can be demonstrated analytically that the SCA approach has a convergence to a local optimal point\cite{sca}. We should apply the following lower bound $\log_2{(1+z)}\leq \alpha \log_2{z}+\beta$, where $\alpha=\frac{\bar{z}}{1+\bar{z}}$, and $\beta=\log_2(1+\bar{z})-\frac{\bar{z}}{1+\bar{z}} \log_2\bar{z} $ in which $\bar{z} \in [0,\infty)$ is a positive real-valued number. As a consequence, the rate of the $k$-th user which is connected to the $m$-th RRH on bandwidth $B$ and $b_{m,z}$ are respectively transformed to the following equations.
 \begin{equation}\label{r1} 
 \hat{r}_{k_t}\geq \alpha_{k,m} \log_2\gamma_{k,m}+\beta_{k,m}, \\ 
 \end{equation} 
  \begin{equation}\label{r2} 
\hat{r}_{k,m,z}\geq\kappa_{k,m,z} \log_2\widehat{\nu}_{k,m,z}+\xi_{k,m,z}, \\  
 \end{equation} 
where $\{\alpha_{k,m}$, $\beta_{k,m}$, $\kappa_{k,m,z}$, $\xi_{k,m,z}\}$ are the approximation constants computed for some $\bar{z}=\gamma_{k,m}$ for (\ref{r1}) and $\bar{z}=\nu_{k,m,z}$ for (\ref{r2}).
By considering the transformation $\bar{p}=\ln{p}$ and $\bar{q}=\ln{q}$, $\bar{E}^{k}_{c}$ and $\bar{P}_T$ are given as 
\begin{align}  \label{ggr}
& \bar{E}^{k}_{c}=\sum_{m=1}^M \frac{-1}{\theta_k } \log \mathbb{E}_{g} \left \{ e^{ \!-\theta_k B T_f (\alpha_{k,m} \log_2(\bar{\gamma}_{k,m})+\beta_{k,m}\! }\right\} \nonumber\\
&\!\! +\! \!\sum_{z=1}^Z\!\! \sum_{m=1}^M \!  \frac{-1}{\theta_k }\!\! \log\! \mathbb{E}_{h}\!\left\{e^{\!\!-\theta_k  \!\left(W_z\!-w_{m,z}\!\right) T_f (\kappa_{k,m,z}\!\! \log_2\!\left(\bar{\nu}_{k,m,z}\!\right)+\xi_{k,m,z})}\!\!\!\right\},\nonumber\\
\end{align} 
where
\begin{align} 
& \bar{\gamma}_{k,m}=\! \frac{ e^{\bar{p}_{k,m}} \mid g_{k,m} \mid ^2}{\sum_{\substack{k'\neq{k} \\  \mid g_{k',m} \mid ^2\leq   \mid g_{k,m} \mid ^2}} e^{\bar{p}_{k',m}} \mid g_{k,m} \mid ^2+B N_0},
\end{align}
\begin{align} 
&\bar{ \nu }_{k,m,z}\!\! =\!\!\frac{e^{\bar{q}_{k,m,z}} \mid h_{k,m,z}\mid ^2}{\sum_{\substack{k'\neq{k} \\ \mid h_{k',m,z}\mid ^2\leq \mid h_{k,m,z}\mid ^2}}\!\!\!\! e^{\bar{q}_{k',m,z}}\! \mid h_{k,m,z} \mid ^2\!\!+\!\! \left(\!W_z\!\!-\!w_{m,z}\! \right)\! N_0}.
\end{align}

\begin{align} \label{ggg}
&\bar{P}_T=\sum_{m=1}^M p_{c,m}+\mathbb{E}_{g}\left\{\sum_{m=1}^M \sum_{k=1}^k  \zeta_m  e^{\bar{p}_{k,m}}\right\}+\sum_{m=1 }^M P_{m,f}\nonumber\\ 
&\hspace{1.2em}+\mathbb{E}_{H}\left\{\sum_{z=1}^Z \sum_{m=1}^M \zeta_m 
\frac{w_{m,z} N_0}{H_{m,z}} (2^{\frac{R_z^{\text{ave-max}}}{T_f w_{m,z}}}-1)\right\}\nonumber \\ 
&\hspace{1.2em}+\mathbb{E}_{h}\left\{\sum_{z=1}^Z \sum_{m=1}^M \sum_{k=1}^K  \zeta_m e^{\bar{q}_{k,m,z}}\right\}. 
\end{align}

In order to be compatible with SCA method, the equality in constraint C8 in problem $\mathcal{P}_2$ should be replaced with an inequality. Therefore, the constraint C8 is replaced by
\begin{equation} 
e^{\bar{p}_{k,m}} e^{\bar{p}_{k,m'}} \leq \varepsilon, ~\forall m,m'\in \mathcal{M}, k\in \mathcal{M}, 
\end{equation} 
where $\varepsilon$ is a small positive number.
Consequently problem $\mathcal{P}_3$ is equivalent to the following problem.
\mathleft
\begin{align} \label{EEprob1} 
\mathcal{P}_3: \max_{\substack{\!\left\{\bar{p}_{k,m}\!\right\}_{k,m},\!\left\{\! \bar{q}_{k,m,z}\!\right\}_{k,m,z}\ , \!\left\{\! w_{m,z}\!\right\}_{m,z} }}  \bar{E}_{c}-\mathbf{q}\bar{P}_T
\end{align}
\text{subject to}
\begin{equation} {
\begin{cases}
\text{C1:}~ \bar{E}^{k}_{c} \geq  R^{\text{ave-max}}_m,~\forall m\in \Psi_z, k ,\nonumber\\
 \text{C4:} ~\mathbb{E}_{g}\!\! \left\{\!\sum_{k=1}^K\!  e^{\bar{p}_{k,m}}\! \right\} +\mathbb{E}_{H} \!\left \{\sum_{z=1}^Z Q_{m,z}\!\right\}\\
 + \mathbb{E}_h\! \! \left\{\!\sum_{z=1}^Z\!\sum_{k=1}^K \! e^{\bar{q}_{k,m,z}}\!\right\}
\leq P^m_{\text{max}},~ \forall m\in \Psi_z, \nonumber\\ 
 \text{C5:}~ e^{\bar{p}_{k,m}} \geq 0,~ \forall  m\in \Psi_z ,k,\nonumber\\ 
 \text{C6:} ~ e^{\bar{q}_{k,m,z}} \geq 0,~ \forall m,z \in \Psi_m ,\nonumber\\ 
 \text{C7:} ~ w_{m,z} \geq 0, ~\forall m,z \in \Psi_m,\nonumber\\ 
\text{C8:} ~ e^{\bar{p}_{k,m}} e^{\bar{p}_{k,m'}} \leq \varepsilon, ~\forall m,m', k. \nonumber
 \end{cases} }
\end{equation} \vspace{\baselineskip}

Presently, the equivalent subtractive problem \eqref{EEprob1} which should be solved at each iteration of Dinkelbach algorithm, is jointly concave with respect to the powers and bandwidth and the duality approach can be applied to find the optimal solutions. The Lagrangian function of the proposed problem is given in \eqref{Lagrange} which is in the top of the next page. 
\begin{figure*}
\begin{align}\label{Lagrange} 
&L(\bar{p}_{k,m},\bar{q}_{k,m,z},w_{m,z}, \mu_{k,m}, \omega_m,\varpi_{k,m,m'})=\bar{E}_{c}-\mathbf{q} \bar{P_T} 
+\sum_{m=1}^M \sum_{k=1}^K  \frac{\mu_{k,m} }{\theta_k} \Big( \log \mathbb{E}_{g}\left \{e^{ -\theta_k B T_f \!\left(\alpha_{k,m} \log_2(\bar{\gamma}_{k,m})+\beta_{k,m} \!\right)}\!\right\}\nonumber\\
&+\sum_{z\in\Psi_m}\!\log\!\mathbb{E}_{h}\!\left\{e^{-\theta_k  (W_z-w_{m,z}\!) T_f \!\left(\kappa_{k,m,z} \log_2\!\left(\bar{\nu}_{k,m,z}\!\right)+\xi_{k,m,z}\!\right)}\!\right\}
-\! R^{\text{ave-max}}_{m} \Big) -\!\!\sum_{m \in \Psi_z} \! \omega_m \Big(\mathbb{E}_{\gamma} \Big\{ \sum_{k=1}^K\! e^{\bar{p}_{k,m}\!\!}\Big\}\!+\mathbb{E}_h \Big\{\sum_{z =1}^Z\!e^{\bar{q}_{k,m,z}} \Big\}\nonumber\\
&+\!\mathbb{E}_{H} \Big\{\sum_{z =1}^Z \frac{w_{m,z}N_0}{H_{m,z}} (2^{\frac{R_z^{\text{ave-max}}}{T_f w_{m,z}}}-1)\Big\}
-P^m_{\text{max}}\Big)\! +\! \sum_{ \substack{m'=1\\ m'\neq m}}^M \sum_{ \substack{m =1\\}}^M \sum_{k=1}^K \varpi_{k,m,m'} e^{\bar{p}_{k,m} \bar{p}_{k,m'}}.
\end{align} 
\noindent\rule{18cm}{0.4pt}
\end{figure*}
In this equation, $\{\mu_{k,m}, \omega_m,\varpi_{k,m,m'} \}$ are the non-negative Lagrange multipliers associated to the above constraints respectively. Thus, the dual problem is given as,
\begin{equation}\label{fff} 
\small
\min_{\substack{\mu_{k,m},\omega_m \\\varpi_{k,m,m'}\! }}\! \max_{\substack{ \bar{p}_{k,m} , \bar{q}_{k,m,z}\\w_{m,z}} }L\left (\bar{p}_{k,m},\bar{q}_{k,m,z}, w_{m,z}, \mu_{k,m}, \omega_m, \varpi_{k,m,m'}\right).  
\end{equation}

By applying the Lagrange dual decomposition, Problem \eqref{fff} is decomposed into a slave problem, which solves the Lagrange function with respect to the transmit powers and bandwidth, for fixed Lagrange multipliers and a master problem updates the Lagrange multipliers using the solution of the slave problem \cite{boyd}. The slave and master problems are solved iteratively, and the process converges to the optimal power and bandwidth allocation for each set of MBS user selection strategy $\Psi$. The slave problem is decomposed into two subproblems to find bandwidth and powers iteratively. The closed form of the power allocation strategies are obtained in \emph{Lemma 2}.\\

\emph{\textbf{Lemma 2:}} Consider Problem \eqref{fff}. For the fixed Lagrange multipliers, given $\Psi$ and bandwidth $w_{m,z}$, the optimal power allocation for $p$ and $q$ are obtained as
\begin{align} \label{obtainp}
&\!\! p_{k,m} = \exp[{\frac{\log{\Theta_{k,m}}+\sigma_{k}(\alpha_{k,m}\!\log{\Gamma_{k,m}}+\beta_{k,m})}{-1-\sigma_{k} \alpha_{k,m}}}],
\end{align}
where \!\!\!
\mathleft
\begin{align}
\Theta_{k,m}=\frac{\delta\ln{2}(\mathbf{q}  \zeta_m +\omega_{m}+\varpi_{k,m,m'} p_{k,m'})}{B T_f \alpha_{k,m}(1+\mu_{k,m})},
\end{align}
\mathleft
\begin{align}
\Gamma_{k,m}=\frac { g_{k,m}}{\sum_{i=1,i\neq{k}}^K p_{i,m} g_{k,m}+B N_0},
\end{align}
with 
$\delta=\mathbb{E}_{g} \{e^{-\theta_k B T_f( \alpha_{k,m} log(\bar{\gamma}_{k,m}) +\beta_{k,m})}\}$ and $\sigma_{k}=\theta_k B T_f$.
\begin{align}\label{obtainq}
& q_{k,m,z}=\exp[{\frac{\log{\Upsilon_{k,m,z}}+\varsigma_{k}(\xi_{k,m,z}\!\log{\Lambda_{k,m,z}}+\kappa_{k,m,z})
}{-1-\varsigma_{k} \xi_{k,m,z}}}],
\end{align}
where 
\mathleft
\begin{align}
\Lambda_{k,m,z}= \frac { h_{k,m}}{\sum_{i=1,i\neq{k}}^K q_{i,m,z} h_{k,m}+\left(W_z-w_{m,z}\right) N_0},
\end{align}
\mathleft
\begin{align}
\Upsilon_{k,m,z}= \frac{\lambda\ln{2}\left(\omega_{m}-\mathbf{q} \zeta_m\right)}{(W_z-w_{m,z}) T_f\xi_{k,m,z} (1+\mu_{k,m})},
\end{align}
with $\lambda=\mathbb{E}_{h} \{e^{-\theta_k (W_z-w_{m,z}) T_f( \xi_{k,m,z} log(\bar{\vartheta}_{k,m,z}) +\kappa_{k,m,z})} \}$ and $\varsigma_{k}=\theta_k \left(W_z-w_{m,z}\right)T_f$.\begin{IEEEproof} See Appendix B.\end{IEEEproof}
It can be seen that the power allocations in (\ref{obtainp}) and (\ref{obtainq}) are different from the classical water-filling solution and it depends on the delay QoS requirements, channel gain and bandwidth.
To obtain the optimal bandwidth allocation for the given power allocation schemes, we set the derivative of the Lagrangian function with respect to $w_{m,z}$ to zero as follows, 
\begin{align} \label{obtainw}
&\frac{\partial {L}}{\partial{w_{m,z}}}\! = \sum_{k=1}^K[ T_f (1+\mu_{k,m})(\xi_{k,m,z}-\kappa_{k,m,z}{\log_2 \nu_{k,m,z}}) \nonumber\\ 
&\hspace{3em}-\frac{T_f\xi_{k,m,z}} {N_0+I_{k,m,z}(W_z-w_{m,z})^{-1} }] -\frac{N_0 (\omega_{m}+\mathbf{q}\zeta_{m})}{H_{m,z}}\nonumber\\
&\hspace{3em}\times (1-2^{\frac{R^{\text{ave-max}}_z}{w_{m,z}}}-\ln 2\frac{R^{\text{ave-max}}_z}{w_{m,z}}2^{\frac{R^{\text{ave-max}}_z}{w_{m,z}}}=0.
\end{align}
The form of \eqref{obtainw} does not allow us to find an elegant closed-form analytic solution for the bandwidth allocation therefore we resort to a numerical search, such as Newton's algorithm. Based on the above derivations, the overall resource allocation algorithm is formulated by alternatively optimizing the power allocation strategies according to \eqref{obtainp} and \eqref{obtainq} for fixed bandwidth, and then optimizing the bandwidth for fixed powers, as described above. \\
~~Next, we turn our attention to solve the master problem. Since the master problem is always convex, the sub-gradient method updates the Lagrange multipliers with guaranteed convergence and as it is mentioned in \cite{yu2006dual}, the gradient algorithm is more suitable for distributed implementation, where each user may update its own dual variable autonomously. By applying the gradient method, which leads to the following update formulas: 
\mathleft
\begin{align}\label{obtain1} 
&\mu^{(t+1)}_{k,m}\!\!=\bigg[\mu^{(t)}_{k,m}+\!\beta_{1}\!\left(\sum_{m=1}^M\sum_{k=1}^K \bar{E}^{k}_{c}-\! R^{ave-max}_{m}\!\right)\bigg]^+,\nonumber\\
\end{align}
\begin{align}
&\omega^{(t+1)}_m=\bigg[\omega^{(t)}_m-\beta_{2}\bigg( P^m_{\text{max}}-\sum_{m=1}^M p_{k,m} -\sum_{z =1}^Z  \sum_{m \in \Psi_z}q_{k,m,z} \nonumber\\ 
&-\mathbb{E}_{H}\!\left\{\sum_{z =1}^Z \sum_{m \in \Psi_z}\frac{(W_z-w_{m,z}) N_0 }{H_{m,z}}(2^{\frac{R_z^{\text{ave-max}}}{T_f\! \left(W_z-w_{m,z}\right)}}-1)\! \right\}\! \bigg)\!\bigg{]}^+\!\!, 
\end{align}
\begin{equation}\label{obtain3} 
\varpi^{(t+1)}_{k,m,m'}\!\!=\bigg[\!\varpi^{(t)}_{k,m,m'}+\!\beta_3 \sum_{\substack{m =1\\m\neq m'}}^M \sum_{m' =1}^M \sum_{k=1}^K\! p_{k,m} p_{k,m'}\bigg]^+,\\
\end{equation}
where $t$ is the iteration index, $\beta_1$, $\beta_2$, and $\beta_3$  are  positive step-sizes for the $t$-th iteration, (e.g., $\frac{1}{\sqrt{t}}$ from \cite{boyd}) and $[.]^+$ denotes the projection onto the non-negative orthant. Now, we turn our attention to obtain MBS user assignment by matching theory.
\subsection{MBS User Assignment by Distributed Stable Matching} 
Consider problem $\mathcal{P}_4$ with the set of RRHs as $\mathcal{M}=\{1,...,M\}$ and set of MBS users as $\mathcal{Z}=\{1,...,Z\}$. Each RRH prefers to serve the $z$-th MBS user which can get more bandwidth from it. At the same time, each MBS user should be connected to the RRH which gives it more power. Clearly, this is a conflict situation, since two RRHs might prefer the same MBS users. To obtain a distributed implementation in the MBS user assignment problem, we would need to implement a rejection mechanism which does not require information transfer to a central point or the use of shared memory. In this context, we intend to use stable matching which provides a framework to match RRHs and MBS users in a stable way, where stable means that no RRH can gain by unilaterally changing its assigned MBS user\cite{alessio1}. In order to make this general idea formally precise, at first we should have some discussion about matching theory.\\
Matching occurs in every aspects of our lives. When there are non-divisible goods and entities with different interests in these goods, there is a corresponding matching market. Based on the local information, each entity has preferences on the goods which is determined by utility function. Let us define RRH as entities and MBS as goods, then the utility function of the $m$-th RRH on the $z$-th MBS user is defined as $u_{z,m}$. Hence, the $z$-th MBS user is acceptable to the $m$-th RRH if it leads to a larger utility value than the other MBS users as $u_{z,m}>u_{z',m}, \forall z' \in Z \setminus \{z\}$. 
It should be emphasized that the MBS users also have a preference relation, which, for the case at hand, is based on the utility value $u_{z,m}$. If the same utility value determines both preference lists (as in our case), an easier representation is to replace the preference relations with the utility matrix $\mathbf{U} =\{u_{z,m}\}_{z,m}$. Finally matching market is completely defined by the tuple $(\mathcal{M},\mathcal{Z},\mathbf{U})$.
The matching problem is  \textit{stable} matching if no RRH/MBS user wants to change the current assignment.

Now, we turn our attention to define the utility function $u_{z,m}$. Due to sum-based form of the total effective capacity with respect to $x_{z,m}$, the objective function can be decoupled and the parts with $x_{z,m}$ can be defined as utility function as follows,
\begin{align}\label{uu}
&u_{z,m}\!\!=\!\!\sum_{k=1}^K\!\!  \left\{\!\!\frac{-\log\! \mathbb{E}_{h}\! \left\{e^{\!\! -\theta_k(W_z\!-w_{m,z}\!) T_f \left(\kappa_{k,m,z}\!\! \log_2\!(\nu_{k,m,z}\!)+\xi_{k,m,z} \!\right)}\right\}}{ \theta_k }\!\! \!\right\}\nonumber\\
&-\mathbf{q}\left(\mathbb{E}_{h}\left\{\zeta_m q_{k,m,z}\right\}+ \mathbb{E}_{H}\left\{\zeta_m \frac{N_0 w_{m,z}}{H_{m,z}}(2^{\frac{R_z^{\text{ave-max}}}{T_f w_{m,z}}}-1)\right\}\right).
\end{align}
By considering stable matching perspectives, the RRH-MBS user assignment problem is formulated by identifying the RRH and MBS user sets $\mathcal{M}$ and $\mathcal{Z}$ as the two groups to be matched, while the preference lists are represented by the utility function \eqref{uu}. Specifically, considering the utility matrix $\mathbf{U}$, sorting the entries of the elements in the $z$-th row of $\mathbf{U}$ in the decreasing order yields the preference list of the $z$-th MBS user, whereas sorting the elements of the $m$-th column of $\mathbf{U}$ in decreasing order yields the preference list of the $k$-th RRH. In general matching problems, each matching group have special matching matrix which represents its preference list. In our case, the preferences of all RRHs and MBS users are uniquely determined by a single matrix instead of two matrices which is the special case of the general matching problems. Matching problems in which all preference lists can be determined by a single matrix enjoy stronger properties than general matching problems. With given utilities $\mathbf{U}$ and $\mathbf{q}$, we define the function $D: (\mathbf{U},\mathbf{q}) \in \mathbb{R}^{M\times Z}\times \mathbb{R} \rightarrow D(\mathbf{U},\mathbf{q})\in \mathcal{A}$ where  $\mathcal{A}$ is the set of all feasible assignment which maps a preference matrix $\mathbf{U}$ and a parameter $\mathbf{q}$ to the corresponding stable matching $D: (\mathbf{U},\mathbf{q})$. Consequently, the overall resource allocation algorithm can be formally stated as in Algorithm 1.\par
\begin{algorithm}[]
{\small
\begin{center}
\normalsize \textbf{Algorithm 1}
\end{center}
\textbf{ Resource Allocation Algorithm}\\
\vspace{2pt}
\noindent\rule{8cm}{0.4pt}\\
1: $q=0$\\
~2: \text{Solve $F(\mathbf{q})=\max_{\mathcal{F}} \left(\bar{E}_{c}-\mathbf{q}
\bar{P}_{T} \right)$.} \\
\noindent\rule{8cm}{0.4pt}\\
\While {$F(\mathbf{q}) \geq \epsilon$}
{3: \text{Compute $p_{k,m}$, $q_{k,m,z}$, $w_{m,z}$ and $\mathbf{U}$ based on Eq. } \\
\text{(\ref{obtainp}), (\ref{obtainq}), (\ref{obtainw}) and  \eqref{uu}.}\\
4:  $\mathbf{x} = D(\mathbf{U},\mathbf{q})$\\
5: \text{Compute $F(\mathbf{q})$ for given $\mathbf{q}$. }\\
6: $\mathbf{q}=\frac{\bar{E}_{c}}{\bar{P}_{T}}$\\
}
\noindent\rule{8cm}{0.4pt}\\
7:Return the optimal solutions $p_{k,m}$,$q_{k,m,z}$, $w_{m,z}$,  and $x_{z,m}$.}
\noindent\rule{8cm}{0.4pt}
\label{tab:MUmatch}
\end{algorithm}

\subsection{Complexity Analysis}
The optimal solution of this problem to select the best MBS user is exhaustive search algorithm. It is an algorithm to solve the discrete and combinatorial optimization problem. It searches over all possible MBS users for each RRH and it requires $\mathcal{O}{(2^MZK)}$ operations, which is prohibitive for large number of $M$. 
Greedy algorithm and matching theory are two promising approaches  having extremely low computational complexity, but weaker optimality properties. The greedy algorithm which is proposed in \cite{QW1} for MBS selection, should be run outside of Dinklebach algorithm and it needs to solve fractional programming at each iteration which leads to complexity $\mathcal{O}{(M^2ZK)}$. However, the matching algorithms execute the preference matrix inside the Dinkelbach algorithm and select the best matched RRH and MBS users based on sorting the preference matrix. Hence, for the case at hand, the stable matching is guaranteed to converge in $M$ steps with complexity $\mathcal{O}(Mlog(M))$ and the overall complexity of Algorithm 1 is obtained as $\mathcal{O}{(ZKMlog(M))}$ which is much lower than the Greedy algorithm. 

\section{ SIMULATION RESULTS} 
In this section, we illustrate the numerical results to verify the analytical results of the proposed delay aware and effective energy efficient resource allocation schemes in the NOMA based H-CRAN. The considered H-CRAN consists of $1$ MBS, $2$ to $12$ RRHs for different scenarios, $2$ MBS user and up tp $14$ users. The MBS is located in the center of the cell area, while the RRHs, MBS users and users are uniformly distributed. The bandwidth of each RRH is $B=700$ kHz and the bandwidth of MBS is $W_{MC}=550$ kHz and the slot duration is $1ms$.  Without loss of generality, we assume that all MBS users have identical parameters, i.e., the same amount of available bandwidth $W_z=200$ kHz and average required rate $R_z^{ave-max}=700$ Kbps and also the average rate of the each fronthaul link is specifies as $R_m^{ave-max}=1000$ Kbps. Channels between all users even the MBS users and RRH are flat fading with Rayleigh distributions and  have been modelled as $\alpha^2 |\frac{d_0}{d_k}|^{(2.5)}$, where $\alpha$ is a standard complex circularly symmetric Gaussian random variable which models fading effects and $|\frac{d_0}{d_k}|$ models the communication power path-loss at $d_0 = 1m$ as the reference distance, and $d_k$ denotes the physical distance between the transmitter and the $k$-th user. The noise power is $-102 dBm$. For each RRH, the reciprocal of the drain efficiency of the power amplifier, circuit power consumption and power consumption of the fronthaul fiber link are given by $\zeta=16\%$, $P_c=0.01 W$ and $P_f=1 W$. 

We also stress that the proposed optimization problem contains statistical expectations as a direct effect of the fact that the long-term performance measure is required to define effective capacity. Since closed form expressions are not available,  numerical computations based on Monte-Carlo simulation with Matlab are obtained.
\subsection{Convergence of the Proposed Iterative Algorithm}
\begin{figure}
  \begin{center}
   \includegraphics[width=9 cm , height=7 cm]{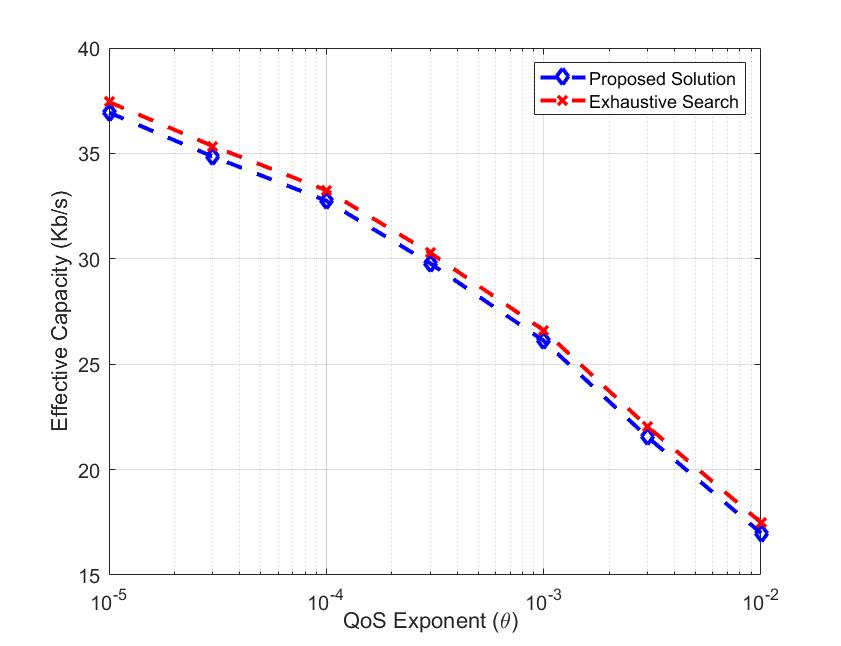} 
    \caption{\footnotesize{Effective Capacity versus QoS exponent ($\theta$) for 3 users and 2 RRHs. }}
    \vspace{-0.5cm}
    \label{11}
  \end{center}
\end{figure}

Fig.\ref{11} shows the total effective capacity for different values of delay QoS exponent for the exhaustive search and our proposed solution. It is obvious that the effective capacity is decreasing function of delay QoS exponent. This is due to the fact that there is a trade-off between achievable low delay and large effective capacity. When the delay requirement is stricter, the effective capacity decreases. Note that when $\theta$ goes to zero, the effective capacity equals Shannon capacity. Moreover, it is observed that the proposed suboptimal solution performs almost as well as the optimal solution with a negligible gap. \par
\begin{figure}
  \begin{center}
   \includegraphics[width=9 cm , height=7 cm]{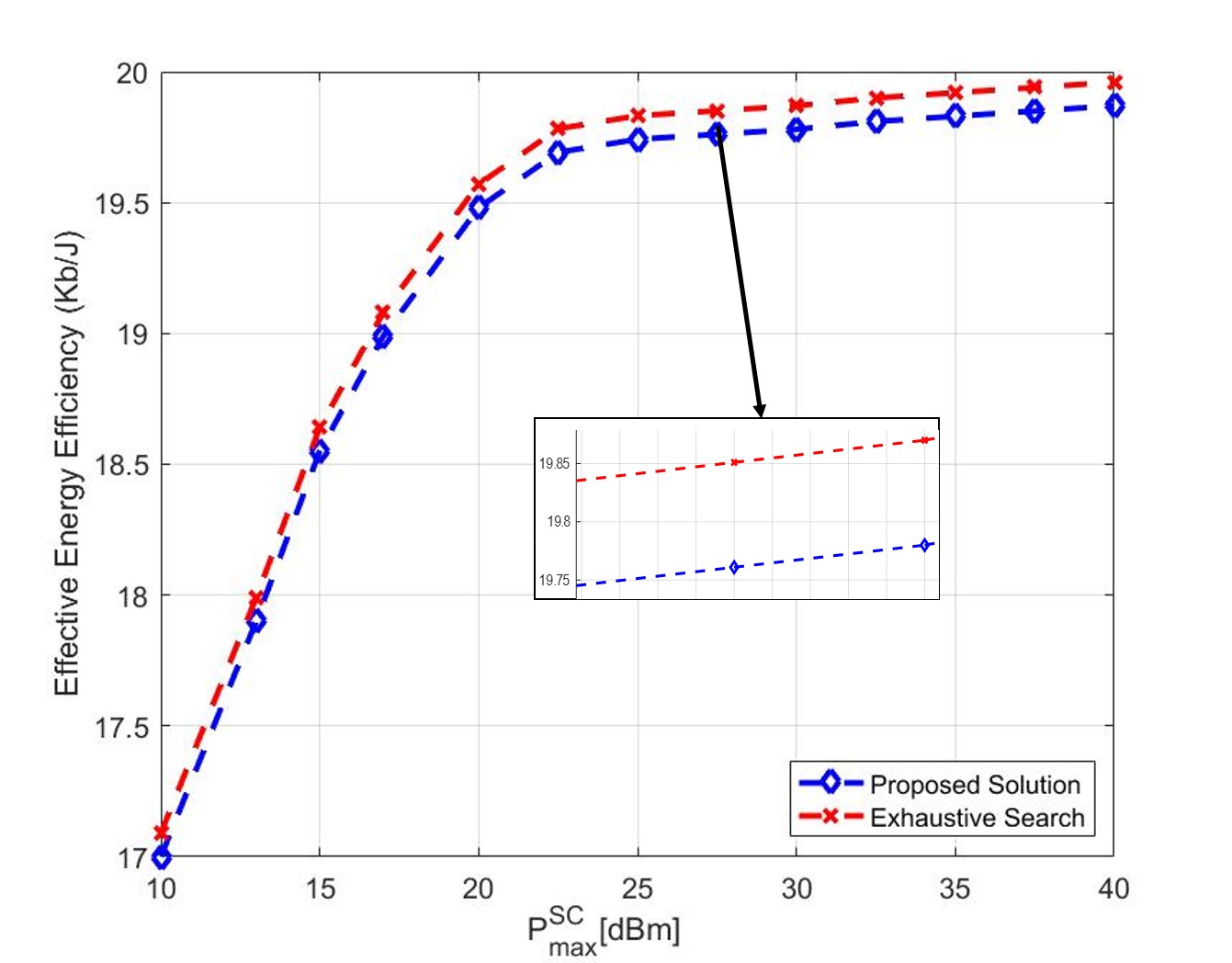} 
    \caption{\footnotesize{Effective Energy Efficiency versus $P_{max}$ for 3 users and 2 RRHs with $\theta=10^{-5}$ }}
    \vspace{-0.5cm}
    \label{165}
  \end{center}
\end{figure}

The comparison between the achieved EEE for the exhaustive search and the proposed algorithm for the different values of  $P^{SC}_{max}$ is given in Fig.\ref{165}. It can be seen that the EEE is an increasing function of $P^{SC}_{max}$. Here we observe that the same as Fig.\ref{11} the proposed suboptimal solution is also performs almost near the optimal solution.  Moreover, both figures demonstrate that our proposed algorithm converges to the near optimal solution. It should be noticed that the simulation results for the different number of users and the RRHs will discuss in further figures with details.\par
\subsection{EEE Performances of the Proposed Solutions}
\begin{figure}
  \begin{center}
   \includegraphics[width=9 cm , height=7 cm]{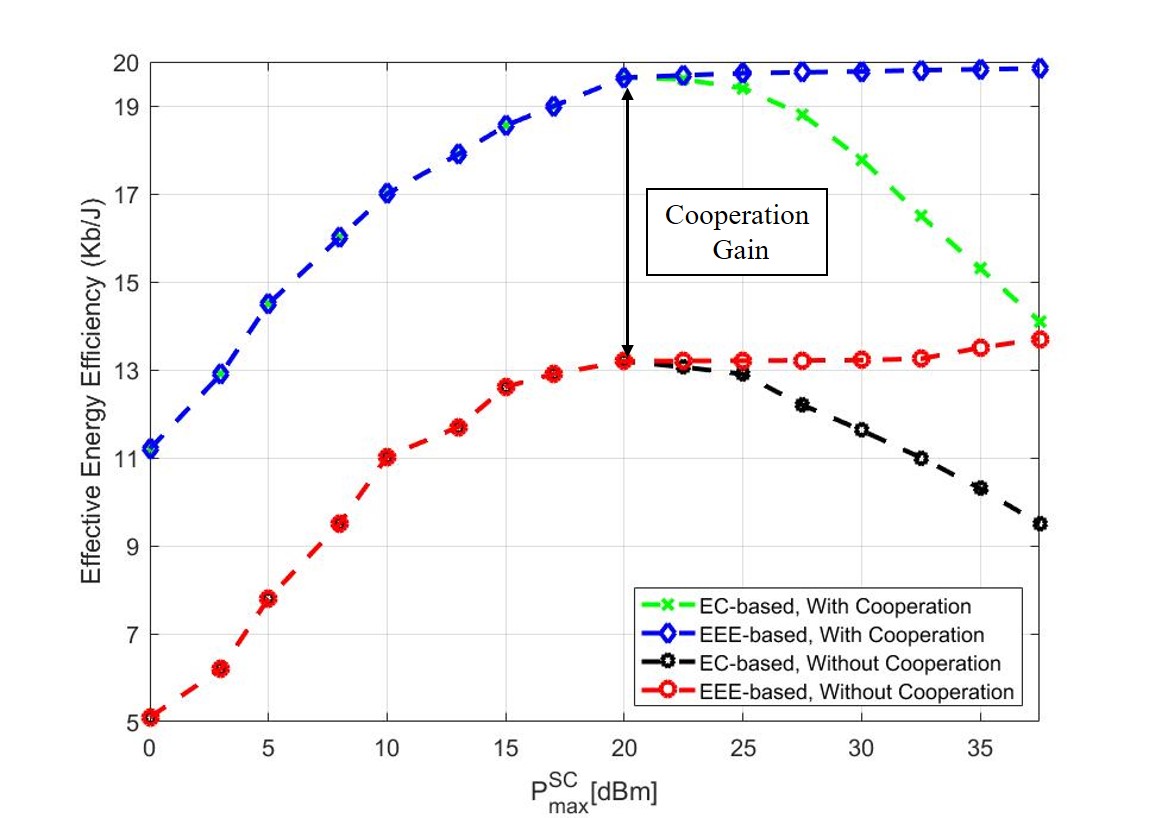} 
    \caption{\footnotesize{Effective Energy Efficiency versus $P_{max}$ for 3 users and 2 RRHs.}}
    \vspace{-0.5cm}
    \label{12}
  \end{center}
\end{figure}

In this section the EEE performance of our proposed algorithm is considered. Fig.\ref{12} shows the variation of the effective energy efficiency obtained by the resource allocations which maximize the effective energy efficiency and the effective capacity versus the maximum average transmit power. In Fig.\ref{12}, we observe that by increasing the maximum available transmit power, the network effective energy efficiency first increases and then saturates when the resources are optimized for the effective energy efficiency maximization. This is due to the fact that the effective energy efficiency is a unimodal function with respect to the transmit powers which can be calculated from the 1st derivative. Specifically when $P^{SC}_{max}$ is so large to get the peak, increasing the transmit power is not useful as it would just decrease the effective energy efficiency. Actually, when the resources are optimized for effective capacity maximization, which is a monotonically increasing function of the transmit power, we see that since the excess transmit power is used to maximize the effective capacity, the attained effective energy efficiency value decreases.

Moreover, Fig.\ref{12} shows that by considering the spectrum and energy cooperation, the effective energy efficiency  in comparison to the traditional architectures is increased which demonstrates the effectiveness of the proposed solution.  
Moreover, it is seen that there is a performance gap between the proposed scheme and the scheme without spectrum and energy cooperation . As we can see that the gap first increases and then reaches a constant. This is because when the transmit power $P^{SC}_{max}$ is limited, RRH does not have sufficient power to serve more MBS users. Hence, our spectrum and energy cooperation is less likely to happen and it limits the improvement in performance. As $P^{SC}_{max}$ increases, RRH has more power to serve its own connected users and more freedom to get more bandwidth from the MBS users. \par
\begin{figure}
  \begin{center}
   \includegraphics[width=9 cm , height=7 cm]{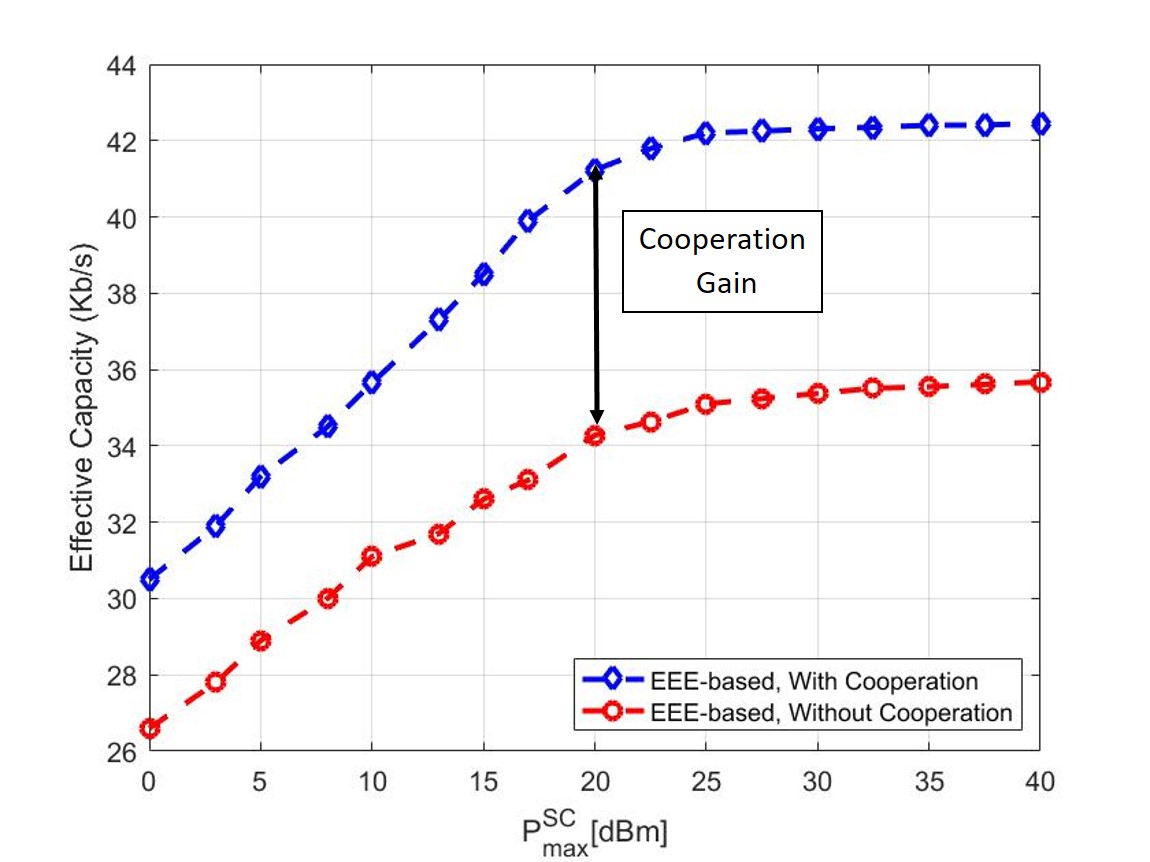} 
    \caption{\footnotesize{Effective Capacity versus $P_{max}$ for 3 users and 2 RRHs. }}
    \vspace{-0.5cm}
    \label{122}
  \end{center}
\end{figure}

Similar conclusions can be drawn from Fig.\ref{122}, which considers a similar scenario as in Fig.\ref{12}, with the effective energy efficiency metric. It can be seen that the effective capacity increases and then saturates when the resources are allocated for effective energy efficiency maximization and also as we can see in the figure, by considering the spectrum and energy cooperation we have about 20\% gain.
\subsection{Performance Comparisons}
\begin{figure}[!htb]
    \begin{subfigure}{\columnwidth}
      \centering
      \includegraphics[width=5.5 cm , height=5 cm]{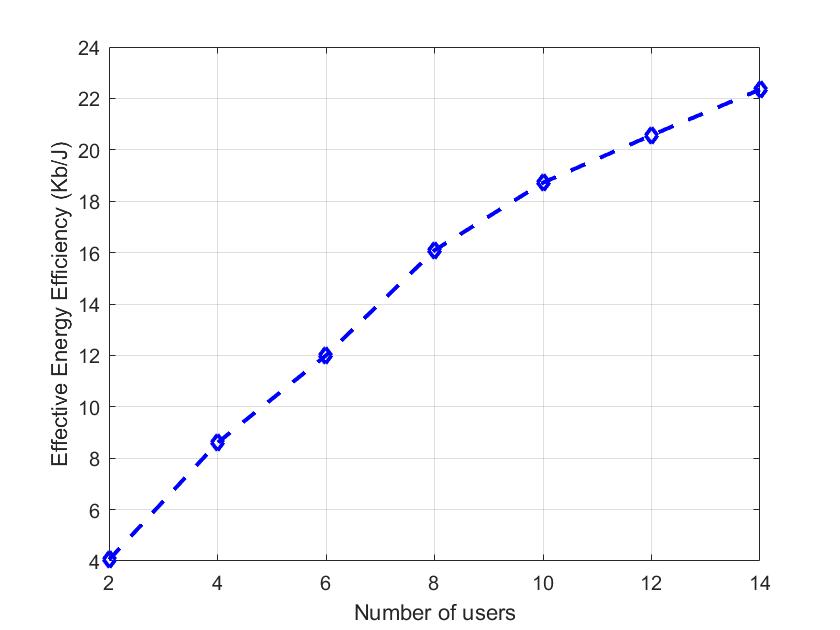}
      \caption{\footnotesize{Effective Energy Efficiency versus the number of users, number of RRH is 10.}}
      \label{13}
    \end{subfigure}
    \begin{subfigure}[b]{\columnwidth}
     \centering
      \includegraphics[width=5.5 cm , height=5 cm]{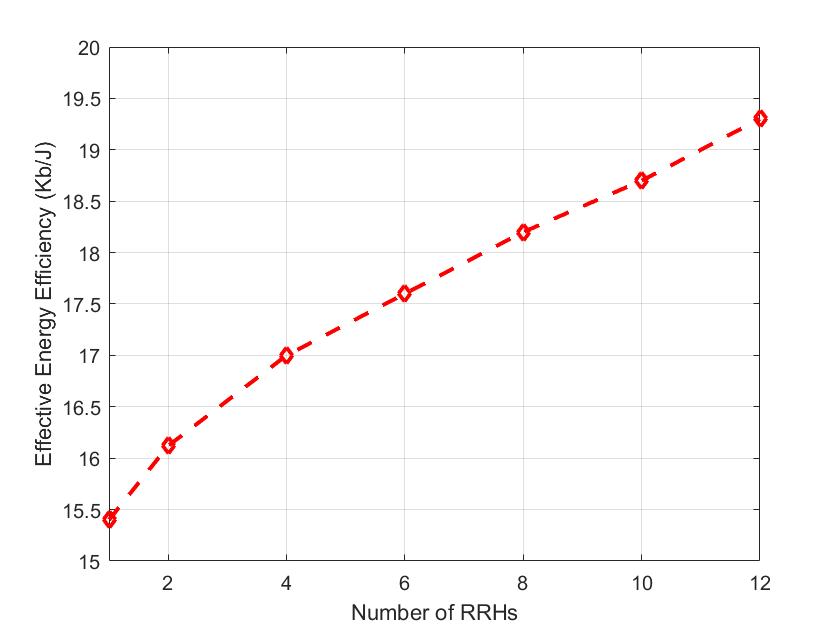}
      \caption{\footnotesize{Effective Energy Efficiency versus the number of RRHs, number of users is 10.}}
      \label{14}
    \end{subfigure}
    \caption{\footnotesize{Impact of different number of users and different number of RRH on EEE and $\theta=10^{-5}$} }
    \label{15}
  \end{figure}

To be applicable in realistic scenarios, the impact of number of users and number of RRHs on effective energy efficiency is considered in this part. In Fig.\ref{13}, the number of RRHs is fixed to 10 and the number of users changes from 2 to 14. In Fig.\ref{14}, the number of users is fixed to 10 and the number of RRHs change from 1 to 12. It is shown that the effective energy efficiency is an increasing function of the number of users and the number of RRHs. It is obvious that by increasing the number of RRHs and users in the network, the proposed spectrum and energy cooperation can have more impact and therefore higher spectral and energy efficiency of the system happens. 

\section{conclusion}
In this paper, we have studied the downlink transmission in a NOMA based H-CRAN with different statistical delay requirements. This has been achieved by introducing and
optimizing the EC and EEE metrics. In order to increase the energy and spectral efficiency, the spectrum and energy cooperation is also applied. We formulate an optimization problem to jointly allocate power, bandwidth, and RRHs. Although the problem is fractional and non-convex, an efficient solution with low complexity based on the  Dinkelbach algorithm ,SCA and stable matching is proposed to obtain the near optimal solution. Through numerical simulations, it is shown that there is a trade off between delay and EEE, and EC is a decreasing function of delay. Throughout the work, it has been assumed that the there is no cooperation between different RRHs, which force the users to connect to just one RRH. Relaxing this assumption appears a relevant topic for the problem formulation as future work.
\bibliographystyle{unsrt}
\bibliography{files2,FracProg}

\begin{thebibliography}{10}

\bibitem{15}
M.~Sinaie, A.~Zappone, J.~Jiang, E.~A. Jorswieck, and P.~Azmi.
\newblock A novel power consumption model for effective energy efficiency in
  wireless networks.
\newblock {\em IEEE Wireless Commun. Lett.}, 5(2):152--155, 2015.

\bibitem{11}
M.~Sinaie, P.~H. Lin, A.~Zappone, P.~Azmi, and E.~A. Jorswieck.
\newblock Delay-aware resource allocation for {5G} wireless networks with
  wireless power transfer.
\newblock {\em IEEE Trans. Veh. Technol.}, 67(7):5841--5855, 2018.

\bibitem{zhang2018energy}
H.~Zhang and et~all.
\newblock Energy-efficient resource allocation in {NOMA} heterogeneous
  networks.
\newblock {\em IEEE Wireless Commun.}, 25(2):48--53, 2018.

\bibitem{QW1}
Q.~Wu, G.~Ye Li, W.~Chen, and D.~Wing~Kwan Ng.
\newblock Energy-efficient small cell with spectrum-power trading.
\newblock {\em IEEE J. Sel. Areas Commun.}, 34(12):3394--3408, 2016.

\bibitem{peng2014energy}
M.~Peng and et~all.
\newblock Energy-efficient resource assignment and power allocation in
  heterogeneous cloud radio access networks.
\newblock {\em IEEE Trans. Veh Technol}, 11(12):5275--5287, 2014.

\bibitem{rbn}
SN. Moosavi and V.~Pourahmadi.
\newblock Opportunistic multiple access (oma) for crowdsensing networks with
  sparse activation.
\newblock {\em Trans Emerg. Telecommun. Technol}, 30(4), 2019.

\bibitem{liu2013green}
D.~Liu, W.~Wang, and W.~Guo.
\newblock Green' cooperative spectrum sharing communication.
\newblock {\em IEEE Commun. Lett}, 17(3):459--462, 2013.

\bibitem{guo2014joint}
Y.~Guo, J.~Xu, L.~Duan, and R.~Zhang.
\newblock Joint energy and spectrum cooperation for cellular communication
  systems.
\newblock {\em IEEE Trans. Commun.}, 62(10):3678--3691, 2014.

\bibitem{xie2012energy}
X.~Renchao, Y.~F. Richard, J.~Hong, and L.~Yi.
\newblock Energy-efficient resource allocation for heterogeneous cognitive
  radio networks with femtocells.
\newblock {\em IEEE Trans. Wireless Technol.}, 11(11):3910--3920, 2012.

\bibitem{cheng2014heterogeneous}
W.~Cheng, X.~Zhang, and H.~Zhang.
\newblock Heterogeneous statistical {QoS} provisioning for downlink
  transmissions over mobile wireless cellular networks.
\newblock {\em IEEE Glob. Commun. Conf. (GLOBECOM)}, pages 4622--4628, 2014.

\bibitem{cheng2016decentralized}
W.~Cheng, X.~Zhang, and H.~Zhang.
\newblock Decentralized heterogeneous statistical {QoS} provisioning for
  uplinks over {5G} wireless networks.
\newblock {\em IEEE Glob. Commun. Conf. (GLOBECOM)}, pages 1--7, 2016.

\bibitem{amjad2019effective}
M.~Amjad, L.~Musavian, and M.~H. Rehmani.
\newblock Effective capacity in wireless networks: A comprehensive survey.
\newblock {\em IEEE Commun. Surveys Tut.}, pages 1--7, 2019.

\bibitem{WU1}
D.~Wu and N.~Rohit.
\newblock Effective capacity-based quality of service measures for wireless
  networks.
\newblock {\em IEEE Commun. Surveys Tut.}, 11(1):91--99, 2006.

\bibitem{chen1}
C.~Mingzhe, W.~Saad, J.~Wang C.~Yin, and M.~Debbah.
\newblock Echo state networks for proactive caching in cloud-based radio access
  networks with mobile users.
\newblock {\em IEEE Trans. Wireless Commun.}, 16(6):3520--3535, 2017.

\bibitem{zhao1}
Z.~Zhongyuan, M.~Peng, Z.~Ding, J.~Wang, W.~Wang, and H.~Vincent Poor.
\newblock Cluster content caching: An energy-efficient approach to improve
  quality of service in cloud radio access networks.
\newblock {\em IEEE J. Sel. Areas Commun}, 34(5):1207--1221, 2016.

\bibitem{Du1}
D.~Qinghe and C.~Zhang.
\newblock Queuing analyses and statistically bounded delay control for two hop
  green wireless relay transmissions.
\newblock {\em Concurrency and Computation: Practice and Experience},
  25(9):1050--1063, 2013.

\bibitem{ismaiel2018analysis}
B.~Ismaiel and all et.
\newblock Analysis of effective capacity and throughput of polling-based
  device-to-device networks.
\newblock {\em IEEE Trans. Veh. Technol.}, 67(9):8656--8666, 2018.

\bibitem{choi2017effective}
J.~Choi.
\newblock Effective capacity of {NOMA} and a suboptimal power control policy
  with delay {QoS}.
\newblock {\em IEEE Trans. Commun.}, 65(4):1849--1858, 2017.

\bibitem{ns}
M.~Sinaie, D.~Wing~Kwan Ng, and E.~A. Jorswieck.
\newblock Resource allocation in {NOMA} virtualized wireless networks under
  statistical delay constraints.
\newblock {\em IEEE Wireless Commun. Lett.}, 7(6):954--957, 2018.

\bibitem{2}
Y.~Wenjuan, L.~Musavian, and Q.~Ni.
\newblock Link-layer capacity of {NOMA} under statistical delay {QoS}
  guarantees.
\newblock {\em IEEE Trans. Commun.}, 66(10):4907--4922, 2018.

\bibitem{concept}
M.~D. Ali, H.~Tabassum, and E.~Hossain.
\newblock Dynamic user clustering and power allocation for uplink and downlink
  non-orthogonal multiple access {NOMA} systems.
\newblock {\em IEEE Access}, 4(3):6325--6343, 2016.

\bibitem{alpha1}
I.~Wong and B.L. Evans.
\newblock Optimal downlink ofdma resource allocation with linear complexity to
  maximize ergodic rates.
\newblock {\em IEEE Trans. Wireless Commun.}, 7(3):962--971, 2008.

\bibitem{sig}
J.~Li and et~all.
\newblock Energy-efficient joint congestion control and resource optimization
  in heterogeneous cloud radio access networks.
\newblock {\em IEEE Trans. Veh Technol}, 65(12):9873--9887, 2016.

\bibitem{e1c}
D.~Wu and R.~Negi.
\newblock Effective capacity: a wireless link model for support of quality of
  service.
\newblock {\em IEEE Trans. Wireless Commun.}, 2(4):630--643, 2003.

\bibitem{tang1}
T.~Jianhua, W.~Peng Tay, and T.~QS Quek.
\newblock Cross-layer resource allocation with elastic service scaling in cloud
  radio access network.
\newblock {\em IEEE Trans. Wireless Commun.}, 14(9):5068--5081, 2015.

\bibitem{wu2005utilizing}
D.~Wu and R.~Negi.
\newblock Utilizing multiuser diversity for efficient support of quality of
  service over a fading channel.
\newblock {\em IEEE Trans. Veh. Technol.}, 54(3):1198--1206, 2005.

\bibitem{42}
J.~P. Crouzeix and J.~A. Ferland.
\newblock Algorithms for generalized fractional programming. mathematical
  programming.
\newblock {\em Mathematical Programming}, 52(1-3):191--207, 1991.

\bibitem{alessio1}
A.~Zappone, E.~Jorswieck, and A.~Leshem.
\newblock Distributed resource allocation for energy efficiency in {MIMO}
  {OFDMA} wireless networks.
\newblock {\em IEEE J. Sel. Areas Commun}, 99(12):53451--3465, 2016.

\bibitem{sca}
J.~Papandriopoulos and J.~S Evans.
\newblock {SCALE}: A low-complexity distributed protocol for spectrum balancing
  in multiuser {DSL} networks.
\newblock {\em IEEE Trans. Inf. Theory}, 55(8):3711--2724, 2009.

\bibitem{boyd}
S.~Boyd and L.~Vandenberghe.
\newblock Convex optimization.
\newblock {\em Cambridge, U.K.: Cambridge Univ. Press}, 2004.

\bibitem{yu2006dual}
Y.~Wei and R.~Lui.
\newblock Dual methods for nonconvex spectrum optimization of multicarrier
  systems.
\newblock {\em IEEE Trans. Commun.}, 54(7):1310--1322, 2006.

\bibitem{sun}
S.~Yan, D.~Wing~Kwan Ng, D.~Zhiguo, and R.~Schober.
\newblock Optimal joint power and subcarrier allocation for {MC-NOMA} systems.
\newblock {\em Glob. Commun. Conf. (GLOBECOM), IEEE}, pages 1--6, 2016.

\bibitem{38}
D.~Tse and V.~Pramod.
\newblock Fundamentals of wireless communications.
\newblock {\em Cambridge, U.K.: Cambridge Univ. Press}, 2005.

\end{thebibliography}
\appendices
\section{ Proof of Lemma 1}
In our NOMA system, we assume that SIC receiver at the $k$-th user wants to cancel the interference from the $k'$-th user. The $k$-th user SIC receiver can decode and remove the interference from the $k'$-th user, if the SINR of the $k'$-th user, which its data stream is decoded at the $k$-th user, is higher than its own SINR \cite{sun}. Therefore, mathematically we have $\gamma_{k,m}(k')>\gamma_{k',m}(k')$, where $\gamma_{k,m}(k')$ is the SINR of the $k'$-th user at the $k$-th user and $\gamma_{k',m}(k')$ is the SINR of the $k'$-th user.
Therefore the SINR condition for the successive decoding is given by the following equation, 
\mathleft
\begin{align} \label{SINRR} 
\!\frac{ p_{k',m} |g_{k,m}|^2} {\mathcal{I}_{k,m} + B N_0} \geq \frac{ p_{k',m} |g_{k',m}|^2}{\mathcal{I}_{k',m} + B N_0}\!, 
\end{align} 
\normalsize
where $\mathcal{I}_{k,m}=\sum_{\tiny{\substack{i\neq{k'}\\ |g_{k',m}|^2\leq |g_{i,m}|^2}}}p_{i,m}|g_{k,m}|^2$ and $\mathcal{I}_{k',m}=\sum_{\tiny{\substack{i\neq{k'}\\ |g_{k',m}|^2\leq |g_{i,m}|^2}}}\! p_{i,m}|g_{k',m}|^2$, then the equation \eqref{SINRR} is equivalent to,
\mathleft
\begin{align} \label{noma3}
&  B N_0\!\!\left(|g_{k',m}|^2-|g_{k,m}|^2\right)\!\! +\!\! |g_{k,m}|^2 |g_{k',m}|^2(\!\!\!\!\!\!\!\sum_{\tiny{\substack{i\neq{k'}\\ |g_{k',m}|^2\leq |g_{i,m}|^2}}}\!\!\!\!\!\! p_{i,m}\!)\nonumber\\
&-(\!\!\!\!\!\!\!\!\sum_{\tiny{\substack{i\neq{k'}\\|g_{k',m}|^2\leq |g_{i,m}|^2}}}\!\!\!\!\!\!p_{i,m})\leq 0 
=B N_0 \left(|g_{k',m}|^2-|g_{k,m}|^2\right) \leq 0, \nonumber\\
 \end{align}
\normalsize
 which results to $|g_{k',m}|^2\leq |g_{k,m}|^2$.\par
 
\section{Proof of Lemma 2}
In the following, we solve the optimal power allocation $p_{k,m}$, by assuming $N$ fading states (or sub-channels) and  introducing an additional index $n$ for each allocated power, namely, $p_{k,m,n}$. For notational ease, we introduce the following definition:
\begin{align}\label{fn}
\mathcal{F}_{n}= e^{-\theta_k B T_f( \alpha_{k,m} log(\bar{\gamma}_{k,m,n}) +\beta_{k,m})},
\end{align}
\mathleft
\begin{align}
&L_1= \sum_{m=1}^M \sum_{k=1}^K (1+\mu_{k,m}) \frac{-1}{\theta_{k}} \log \!\left( \frac{1}{N} \sum_{n=1}^N \mathcal{F}_{n} \!\right)\!\nonumber\\
 &\hspace{1em} - \frac{1}{N} \sum_{n=1}^N {\sum_{m=1}^M \sum_{k=1}^k e^{\bar{p}_{k,m,n}} \left ( \mathbf{q}\zeta_m-\omega_{m}\right) } \nonumber\\
 &\hspace{1em}+\sum_{\substack{m =1\\m\neq m'}}^M \sum_{m' =1}^M \sum_{k=1}^K\ \varpi_{k,m,m'} e^{p_{k,m} p_{k,m'}},
\end{align} 
where $L_1$ corresponds to the terms with $p_{k,m,n}$ in (\ref{Lagrange}). Taking the first derivative of $L_1$ with respect to $p_{k,m,n}$ is as follows:
\mathleft
\begin{align}
&\frac{\partial {L_1}}{\partial{p_{k,m,n}}}=\frac{(1+\mu_{k,m})}{-\theta_k  \bar{\delta}\ln{2}}\mathcal{F'}_n-e^{\bar{p}_{k,m,n}}(\omega_{m} -\mathbf{q}\zeta_m \nonumber\\
&\hspace{3.7em}+\varpi_{k,m,m'} e^{p_{k,m',n}}),
\end{align}
where $\bar{\delta}=\frac{1}{N} \sum_{n=1}^N \mathcal{F}_{n} $. By \eqref{fn} the first derivative of $\mathcal{F'}_{n},$ with respect to $p_{k,m,n}$ is as follows,
\begin{align}
\mathcal{F'}_{n}= -\theta_k B T_f\alpha_{k,m}\mathcal{F}_{n}.
\end{align}
Then set $L_1$ to zero leads to a stationary condition:
\mathleft
\begin{align}
&\frac{\partial {L_1}}{\partial{p_{k,m,n}}}=  \frac{B T_f\alpha_{k,m} (1+\mu_{k,m})}{\bar{\delta}\ln{2}} \mathcal{F}_{n}- e^{\bar{p}_{k,m,n}}(\omega_{m}\nonumber\\
&\hspace{3.7em} -\mathbf{q}  \zeta_m +\varpi_{k,m,m'} e^{p_{k,m',n}})=0,
\end{align}
which can be simply rewritten as 
\begin{align}
\frac{\mathcal{F}_{n}}{e^{\bar{p}_{k,m,n}}}\! =\! \frac{( \omega_{m}-\mathbf{q} \zeta_m+\varpi_{k,m,m'} e^{\bar{p}_{k,m',n}})\bar{\delta}\ln{2}}{B T_f\alpha_{k,m} (1+\mu_{k,m}) }.
\end{align}
Then we can define
\begin{align}
e^{-\sigma_{k} (\alpha_{k,m} \log{(e^{\bar{p}_{k,m,n}} \Gamma}_{k,m,n}) +\beta_{k,m})}\! =\Theta_{k,m,n} e^{\bar{p}_{k,m,n}},
\end{align}
where $\sigma_{k}=\theta_k B T_f$ and $\Theta_{k,m,n}$ and $\Gamma_{k,m,n}$ are given as
\begin{align}
&\Theta_{k,m,n}= \frac{\bar{\delta}\ln{2}(\omega_{m}-\mathbf{q}\zeta_m+\varpi_{k,m,m'} p_{k,m',n})  }{B T_f \alpha_{k,m}(1+\mu_{k,m})},\\
&\Gamma_{k,m,n}= \frac { |g_{k,m,n}|^2}{\sum_{i\neq{k}}^K e^{\bar{p}_{i,m,n}} |g_{k,m,n}|^2+B N_0},
\end{align}
Finally, the optimal power is obtained as follows
\begin{align}
&\!\! p_{k,m,n}\!\! =\!\! \exp[{\frac{\log{\Theta_{k,m,n}}+\sigma_{k}(\alpha_{k,m}\!\log{\Gamma_{k,m,n}}+\beta_{k,m})
}{-1-\sigma_{k} \alpha_{k,m}}}].
\end{align}
Similar to \cite{38}, we let $N \rightarrow \infty $, which yields to
\begin{align}
&\!\! p_{k,m}\!\! =\!\! \exp[{\frac{\log{\Theta_{k,m}}+ \sigma_{k}(\alpha_{k,m}\!\log{\Gamma_{k,m}}+\beta_{k,m})
}{-1-\sigma_{k} \alpha_{k,m}}}],
\end{align}
and $\bar{\delta} \rightarrow \delta $.\\
The proof to find $q_{k,m,z,n}$ follows similar steps as to find $p_{k,m,n}$.  Therefore we introduce the following definition
in (\ref{gn}) 
\mathleft
\begin{align}\label{gn}
\mathcal{G}_{n}= e^{-\theta_k (W_z-w_{m,z}) T_f( \xi_{k,m,z} log(\bar{\vartheta}_{k,m,z,n}) +\kappa_{k,m,z})},
\end{align}
\mathleft
\begin{align}
&\!\!\!\!\!\!\!\!\! L_2=\sum_{z=1}^Z \sum_{m=1}^M \sum_{k=1}^K (1+\mu_{k,m}) \frac{-1}{\theta_{k}} \log \!\left( \frac{1}{N} \sum_{n=1}^N \mathcal{G}_{n} \!\right)\!\nonumber\\
 &\!\!- \frac{1}{N} \sum_{n=1}^N {\sum_{m=1}^M \sum_{k=1}^k e^{\bar{q}_{k,m,z,n}} \left ( \mathbf{q}\zeta_m-\omega_{m}\right)}, 
\end{align} 
where $L_2$ corresponds to the terms with $q_{k,m,z,n}$ in (\ref{Lagrange}). Taking the first derivative of $L_2$ with respect to $q_{k,m,z,n}$ is as follows
\mathleft
\begin{align}
&\frac{\partial {L_2}}{\partial{q_{k,m,z,n}}}=\frac{-(1+\mu_{k,m})}{\theta_k \bar{\lambda} \ln{2}} \mathcal{G'}_{n}- e^{\bar{q}_{k,m,z,n}} \left(\omega_{m}-\mathbf{q}  \zeta_m \right),
\end{align}
where $\bar{\lambda}=\frac{1}{N} \sum_{n=1}^N \mathcal{G}_{n} $. By \eqref{gn} the first derivative of $\mathcal{G'}_{n},$ with respect to $q_{k,m,z,n}$ is as follows
\begin{align}
\mathcal{G'}_{n}= -\theta_k (W_z-w_{m,z}) T_f\xi_{k,m,z} \mathcal{G}_{n}.
\end{align}
Then set $L_2$ to zero leads to a stationary condition
\mathleft
\begin{align}
&\frac{\partial {L_2}}{\partial{q_{k,m,z,n}}}=  \frac{(W_z-w_{m,z}) T_f\xi_{k,m,z} (1+\mu_{k,m})}{\bar{\lambda}\ln{2}} \mathcal{G}_{n} -  \nonumber\\
&\hspace{4.5em}e^{\bar{q}_{k,m,z,n}} ( \omega_{m}-\mathbf{q}  \zeta_m)=0,
\end{align}
which can be simply rewritten as 
\begin{align}
\frac{\mathcal{G}_{n}}{e^{\bar{p}_{k,m,n}}}\! =\! \frac{( \omega_{m}-\mathbf{q} \zeta_m)\bar{\lambda}\ln{2}}{(W_z-w_{m,z}) T_f\xi_{k,m,z} (1+\mu_{k,m}) }.
\end{align}
Then we can define
\begin{align}
e^{-\varsigma_{k}(\xi_{k,m,z} log(\bar{\vartheta}_{k,m,z}) +\kappa_{k,m,z})}\! =\Upsilon_{k,m,z,n} e^{\bar{q}_{k,m,z,n}},
\end{align}
where $\varsigma_{k}=\theta_k \left(W_z-w_{m,z}\right)T_f$ and $\Upsilon_{k,m,z,n}$ and $\Lambda_{k,m,z,n}$ are as follows
\begin{align}
 &\Upsilon_{k,m,z,n}= \frac{\bar{\lambda}\ln{2}\left(\omega_{m}-\mathbf{q} \zeta_m\right)}{(W_z-w_{m,z}) T_f\xi_{k,m,z} (1+\mu_{k,m})},\\
&\Lambda_{k,m,z,n}= \frac { |h_{k,m,z,n}|^2}{\sum_{i\neq{k}}^K q_{i,m,z,n} |h_{k,m,z,n}|^2+\left(W_z-w_{m,z}\right) N_0},
\end{align}
We finally obtain
\begin{align}
&\!\! q_{k,m,z,n}\!\! =\nonumber\\
& \exp[{\frac{\log{\Upsilon_{k,m,z,n}}+\varsigma_{k}\!(\xi_{k,m,z}\!\log{\Lambda_{k,m,z,n}}+\kappa_{k,m,z}\!)
}{-1-\varsigma_{k} \xi_{k,m,z}}}].
\end{align}
Similar to \cite{38}, we let $N \rightarrow \infty $, which yields to (\ref{obtainq}).\\
\begin{align}
&\!\! q_{k,m,z}\!\! = \exp[ {\frac{\log{\Upsilon_{k,m,z}}+\varsigma_{k}\!(\xi_{k,m,z}\!\log{\Lambda_{k,m,z}}+\kappa_{k,m,z}\!)
}{-1-\varsigma_{k} \xi_{k,m,z}}}],
\end{align}
and $\bar{\lambda} \rightarrow \lambda $.
\clearpage

\end{document}